\documentclass[10pt]{JHEP3}

\usepackage{amssymb,amsmath}
\usepackage{graphics}
\usepackage{epsfig}
\usepackage{amsfonts}

\usepackage{psfrag}

\usepackage{amscd}

\newcommand{\be}[1]{ \begin{equation}\label{#1} }
\newcommand{\ee}{\end{equation}}
\newcommand{\bea}[1]{\begin{eqnarray}\label{#1} }
\newcommand{\eea}{\end{eqnarray}}
\newcommand{\eq}[1]{(\ref{#1})}

\newcommand{\sn}{{\rm sn}}
\newcommand{\cn}{{\rm cn}}
\newcommand{\dn}{{\rm dn}}

\newcommand{\g}{\gamma}
\newcommand{\ep}{\epsilon}

\def\s{\sigma}
\def\e{\epsilon}

\def\half{\frac{1}{2}}
\def\d{\partial}
\def\a{\alpha}
\def\b{\beta}

\newcommand{\idn}{{1\relax{\kern-.35em}1}}
\newcommand{\Cf}{\mathbb{C}}

\newcommand{\Zf}{\mathbb{Z}}

\newcommand{\abs}[1]{\left\vert#1\right\vert}

\DeclareMathOperator{\tr}{Tr}

\def\r{{\rm{Re}{[\eta]}}}
\def\i{{\rm{Im}{[\eta]}}}

\preprint{ WIS/03/07-MAR-DPP\\ SLAC-PUB-12405 }
\title{Comments on worldsheet theories dual to free large N gauge theories}
\author{Ofer Aharony$^{1,2}$, Justin R. David$^3$, Rajesh Gopakumar$^3$, Zohar Komargodski$^1$,
Shlomo S. Razamat$^4$\\
$^1$ Department of Particle Physics, Weizmann Institute of
Science,
Rehovot 76100, Israel\\
$^2$ SITP, Department of Physics and SLAC, Stanford University,
Stanford, CA 94305, USA\\
$^3$ Harish-Chandra Research Institute, Chhatnag Road, Jhunsi,
Allahabad 211019, India\\
$^4$ Department of Physics, Technion, Israel Institute of
Technology,
Haifa 32000, Israel\\
E-mails: \email{Ofer.Aharony@weizmann.ac.il},
\email{justin@hri.res.in}, \email{gopakumr@hri.res.in},
\email{ZKomargo@weizmann.ac.il},
\email{razamat@physics.technion.ac.il} }

\abstract{We continue to investigate properties of the worldsheet
conformal field theories (CFTs) which are conjectured to be dual
to free large $N$ gauge theories, using the mapping of Feynman
diagrams to the worldsheet suggested in \cite{Gopakumars}. The
modular invariance of these CFTs is shown to be built into the
formalism. We show that correlation functions in these CFTs which
are localized on subspaces of the moduli space may be interpreted
as delta-function distributions, and that this can be consistent
with a local worldsheet description given some constraints on the
operator product expansion coefficients. We illustrate these
features by a detailed analysis of a specific four-point function
diagram. To reliably compute this correlator we use a novel
perturbation scheme which involves an expansion in the large
dimension of some operators. }

\begin{document}

\section{Introduction}

Since the seminal work of 't Hooft \cite{'tHooft:1973jz} it has been widely believed that large $N$ $SU(N)$
gauge theories (with adjoint matter fields) should have a dual description in terms of closed strings with a
string coupling constant $g_s \sim 1/N$. The original argument for this duality was based on the
reinterpretation of Feynman diagrams as string theory diagrams. Feynman diagrams drawn in 't Hooft's double-line
notation resemble two dimensional surfaces with holes. It was conjectured that there should be an equivalent
description in which the holes get filled up, leading to closed Riemann surfaces without boundaries.

The arguments of 't Hooft  do not give a prescription to construct
the string theory dual to a specific large $N$ gauge theory.
Numerous attempts have been made to directly construct string
theory duals for given field theories. However, success was mainly
achieved when the field theory had a topological interpretation
(Chern-Simons theory \cite{GopakumarVafa} and the Kontsevich model
\cite{Konts} are good examples of this) or the putative dual
string theory was exactly solvable ($d\leq2$ string/ matrix model
duality). This situation has changed following the AdS/CFT
correspondence \cite{Maldacena:1997re}. By now, there are many
examples in which it is known how to find the closed string dual
of gauge theories which can be realized as the world-volume
theories of D-branes in some decoupling limit. In these cases the
closed string dual turns out to be a standard closed string
theory, living in a warped higher dimensional space. In some
cases, for which the gauge theory is strongly coupled, the dual
string background is weakly curved and a gravity approximation of
the string theory may be used. In general (and, in particular, for
all weakly coupled gauge theories), this is not the case, and the
dual string theory is complicated (and does not necessarily have a
geometrical interpretation).

It is interesting to ask is what is the string theory dual of the
simplest large $N$ gauge theory, the free gauge
theory.\footnote{Note that the free limit of an $SU(N)$ gauge
theory is not the same as a theory of $(N^2-1)$ free fields, since
the Gauss law constraint is still imposed on physical states.
Equivalently, we only consider gauge-invariant operators.} There
have been various proposals for how to study the string dual of
free large $N$ gauge theories (see, for instance,
\cite{Gopakumars,FreeFields,StringBits,Joe,Karch,Bonelli:2004ve,
Itzhaki:2004te,Bianchi, Akhmedov,Carfora:2006nj}). It is clear
that the dual string theories must live in a highly-curved
background, which may or may not have a geometrical interpretation
(for four dimensional free gauge theories with massless matter
fields, which are conformally invariant, one expects that any
geometrical interpretation should include an AdS$_5$ factor).

In this paper we continue our study of a specific proposal
\cite{Gopakumars}\footnote{See
\cite{Furuuchi:2005qm,Aharony:2006th,David:2006qc,Yaakov:2006ce}
 for further work on this proposal.} for how to map the Feynman diagrams to
worldsheets. This proposal is based on rewriting the propagators in the Feynman diagrams as integrals over
Schwinger parameters, and mapping these parameters to the moduli of a Riemann surface with holes (which include
the moduli of the closed Riemann surface and the circumferences of the holes).\footnote{The holes here are not
those of the gauge theory Feynman diagram but rather of its dual. They are therefore as many in number as the
vertices of the original graph.} One can then integrate over the parameters of the holes, and translate any
Feynman diagram to a correlation function on the closed string worldsheet. This proposal makes the procedure
advocated by 't Hooft manifest. Most of our discussion will be general, but whenever we need a concrete example
we will consider operators involving adjoint scalar fields in a four dimensional gauge theory.

The mapping of \cite{Gopakumars} gives a closed string theory whose integrated correlation functions (of
physical vertex operators), by construction, reproduce the space-time correlation functions. The worldsheet
theory is also automatically conformally invariant (so that it can be interpreted as a closed string theory in
conformal gauge) and modular invariant. However, the construction does not give a Lagrangian for the worldsheet
theory, and it is not clear from the construction if this worldsheet theory is a standard local conformal field
theory or not.

It was noted in \cite{Aharony:2006th} that the prescription of
\cite{Gopakumars} gives rise to an interesting feature, which
naively is in contradiction to having a well behaved worldsheet
field theory: some of the putative worldsheet correlators localize
on lower dimension subspaces of the moduli space of marked Riemann
surfaces. In a usual local field theory one expects the worldsheet
correlator to be a smooth function on the moduli space, which
naively rules out any localization on subspaces. In this paper we
claim that such a localization can arise in specific field
theories by a special conspiracy between the operator product
expansion (OPE) coefficients. We also discuss the manifestation of
modular invariance in the prescription of \cite{Gopakumars}.
Although modular invariance is guaranteed by the prescription, we
will illustrate that in some amplitudes it is realized in a quite
intricate manner.

This paper is organized as follows. In  \S \ref{Genfeat} we
present the problems of modular invariance and the localization on
the worldsheet and our suggested resolution of these problems. In
\S \ref{Broomlimits} we concentrate on a specific four-point
function : the ``Broom'' diagram. This diagram does not localize
on a subspace of the moduli space, but it has limits in which it
goes over to a localized diagram. We discuss the limits in which
this diagram can illustrate different general features of the
prescription, and we perform an explicit perturbative analysis of
the diagram around exactly solvable limits. In \S
\ref{exactsolutions} we present results about more general exactly
solvable subspaces of the Broom diagram, which serve as a check on
the perturbative analysis. In \S \ref{Fieldtheory} we discuss the
delocalization of the worldsheet amplitudes as illustrated by the
field theory analysis of Broom correlators. We end in \S
\ref{summary and discussions} with a summary and a discussion of
some open problems. In appendix \ref{pertappendix} we discuss the
elliptic function approach to the Strebel problem which arises in
computing the Broom diagram in more detail, and present the
matching between perturbation theory and the exact results. In
appendix \ref{squareapp} two other diagrams which also localize on
a subspace of the moduli space, the Square diagram and the Whale
diagram, are analyzed. We show that the Whale diagram localizes in
a different way than the other diagrams we discuss: it localizes
on a two dimensional subspace of the two dimensional moduli space.

\section{General features}
\label{Genfeat}

In \cite{Gopakumars} a specific prescription was suggested for
mapping the correlation functions of free large $N$ gauge theories
to a string worldsheet, in the 't Hooft large $N$ limit
\cite{'tHooft:1973jz}. This prescription involves rewriting each
Feynman diagram contributing to an $n$-point correlation function
as an integral over the Schwinger parameters of the propagators,
after reduction to a ``skeleton graph'' in which homotopically
equivalent propagators (when the Feynman diagram is interpreted as
a genus $g$ Riemann surface using the double-line notation) are
joined together, and then mapping the space of these Schwinger
parameters to the ``decorated moduli space'' ${\mathcal
M}_{g,n}\times \mathbb R_+^n$. This is the moduli space of  genus
$g$ Riemann surfaces with $n$ marked points, together with a
positive number $p_i$ associated with each point. As described in
detail in \cite{Gopakumars}, this mapping uses the properties of
Strebel differentials\footnote{See
\cite{K.Strebel:1984,Mulase:98,Zvonkine:2002} for details about
Strebel differentials and
\cite{Zwiebach,Moeller:2004yy,Ashok:2006du} for additional
applications of these differentials in string theory.}
 on genus $g$ Riemann surfaces.
After integrating over the $p_i$, this procedure gives a specific
worldsheet $n$-point correlation function associated with this
Feynman diagram which, by construction, reproduces the correct
$n$-point space-time correlation function (upon integration over
the worldsheet moduli).

Certain properties of the mapping proposed in \cite{Gopakumars}
and of the worldsheet correlation functions it leads to can be
understood by general considerations. Two of these properties --
the lack of special conformal invariance in space-time and the
localization of certain amplitudes on lower dimension subspaces of
the moduli space -- were elucidated in \cite{Aharony:2006th}. We
begin this section by discussing another general property of the
resulting worldsheet correlation functions, which is modular
invariance. We then discuss in more detail the interpretation of
the amplitudes which are localized on the moduli space. We argue
that these amplitudes should be interpreted as delta-function
distributions, and we discuss why the appearance of such
distributions in correlation functions is not necessarily in
contradiction with a local worldsheet interpretation. In fact, we
derive some general constraints on the OPE coefficients of the
putative local worldsheet theory which are needed for reproducing
the localized correlation function.

In order to establish an exact sense in which the localized
correlators should be treated as delta-functions, we analyze a
small deformation of the localized diagram. This deformation
involves taking the limit of correlation functions with a large
number of field contractions between the vertex operators of the
original localized diagram, and a small number of additional
contractions. We argue that such correlation functions are smooth
but tend to localize near subspaces of the moduli space, and that
the mapping (and the corresponding Strebel differentials) may be
computed perturbatively in the inverse number of fields (or in the
distance from the localized limit). By using this expansion we can
show that in the limit in which the correlation functions localize
they become delta-function distributions.

In the subsequent sections of the paper, we discuss in detail a particular 4-point function diagram, which we
call the Broom diagram, and we use this diagram to illustrate the general features described in this section. In
particular, the expansion method described in \S \ref{expand} may be used to compute this diagram in some
regions of its parameters.

\subsection{Modular invariance of worldsheet amplitudes defined by
the procedure of \cite{Gopakumars}}\label{modular invariance section}

In this subsection we discuss how modular invariance is realized
 in the prescription suggested in \cite{Gopakumars}. The mapping
 from the Schwinger parameters on the gauge theory side to the
 decorated moduli space of the Riemann surfaces ${\mathcal
 M}_{g,n}\times \mathbb R_+^n$ is, by construction, consistent with the
 action of the modular group on this space.  The mapping uses the
 Strebel theorem to relate the two, which ensures that (for generic
 Feynman diagrams) we cover the whole decorated moduli space (after
 the identifications imposed by the modular group) and we
 cover it exactly once.

However, the way that modular invariance is technically realized can
 in some cases be non-trivial. This is because the mapping of
 \cite{Gopakumars} involves taking the square root of the Strebel
 differential. In cases where this square root has branch cuts,
 one must be careful to choose specific branches in order to ensure
 that the final worldsheet
 correlators are single-valued on the moduli space.

 Several correlators were analyzed in
 \cite{Aharony:2006th,David:2006qc} where it was shown by direct
 computation that these amplitudes are consistent with the modular
 invariance. However, in all of these cases
 the square roots of the Strebel differentials
corresponding to the
 diagrams contributing to those amplitudes did not include fractional
 powers and thus did not have any branch cuts.
In this paper we will analyze in detail a diagram which does have branch cuts for the square root of the Strebel
differential. Note that the existence of the
 branch cuts does not affect ``large modular transformations'' such as $\eta\to
 1/\eta$, but periodicity issues, such as taking two points in a closed
path around each other or taking the torus modular
 parameter $\tau$ to $\tau+1$, should be addressed with more care.  We
 will illustrate the general issue in this subsection by looking at
 general properties of $4$-point functions on the sphere. A specific
 example will be discussed in the next section.

Consider a four-point correlator on the gauge
theory side,
\begin{equation}
\langle {\cal O}_1(x_1) {\cal O}_2(x_2){\cal O}_3(x_3){\cal O}_4(x_4)\rangle.
\end{equation}
In the suggested mapping, we map the space-time operators ${\cal
O}_i(x_i)$ to worldsheet operators $V_{i,x_i}(z)$, and we are
given a specific procedure to calculate
\begin{equation}
\label{wsfour} \langle V_{1,x_1}(0)V_{2,x_2}(1)V_{3,x_3}(\eta) V_{4,x_4}(\infty)\rangle.
\end{equation}
Note that when we write the field theory graph and assign specific
operators to each node, this by itself does not fix the assignment
of the positions of the dual operators on the worldsheet. The
mapping of \cite{Gopakumars} is modular invariant; however, it is
usually convenient to fix the modular freedom by choosing three
specific operators to map to three specific locations, as
indicated in \eqref{wsfour}. Of course, we could also choose to
fix the modular group in any other way, and all such ways are
related by $SL(2,\mathbb{C})$ modular transformations.

Once we choose such a specific fixing of the modular group we can
find the correlator \eqref{wsfour} using the mapping of
\cite{Gopakumars}. However, when we do this we can find that the
relation between the Schwinger parameters and $\eta$ is such that
there are  apparently branch cuts on the $\eta$-plane. This would
lead to an ambiguity in rotating $\eta$ by a full circle around
(say) $z=0$. Since the mapping of \cite{Gopakumars} is
well-defined,\footnote{Given a set of Schwinger parameters there
is a unique set of circumferences and the position on the moduli
space is unique up to the modular group, which we fixed as
described above. In other words, the prescription does not
distinguish between $\eta$ and $\eta e^{2\pi i k}$.} it is clear
that there exists a choice of the branch which leads to an answer
which is single valued, i.e. periodic under a rotation taking
$\eta \to e^{2\pi i} \eta$. However, in some cases this choice of
branch is not the most natural one, and it requires changing the
branch of the square root at some point when we rotate. Obviously,
such a change in the branch is only consistent if the correlator
vanishes at some point on every path taking $\eta \to e^{2\pi i}
\eta$, and this is indeed what we will find in our example in \S
\ref{pertY}.

 \begin{figure}[htbp]
\begin{center}
\epsfig{file=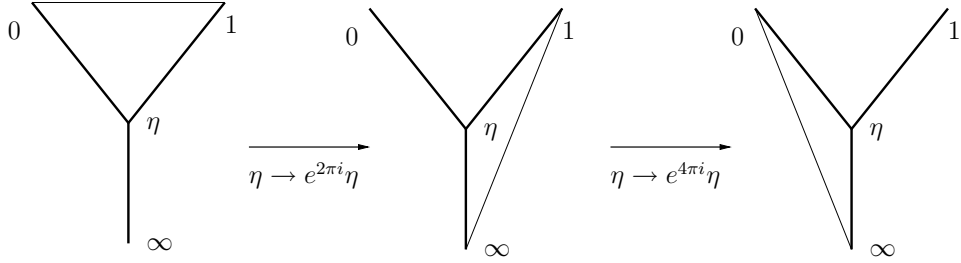, width=5.0in} \caption{The skeleton graph which we refer to as the ``Broom diagram''.
We depict here what happens if we make the {\bf naive} choice of branches, in which we do not change branches
when we rotate $\eta$. We see that for this choice the rotation in the $\eta$ plane changes the position of one
of the edges, and thus changes the correlator we are computing.}\label{Y_spade}
\end{center}
\end{figure}

To illustrate this subtle point consider the ``Broom diagram'',
illustrated in figure \ref{Y_spade}. In this example it turns out,
as we will discuss in more detail in \S \ref{pertY} below, that if
we stay on the same branch of the square root of the Strebel
differential, the topology of the critical graph of this
differential changes as we take $\eta \to e^{2\pi i} \eta$,
leading to a different Feynman diagram, as described in figure
\ref{Y_spade}. Thus, one should be careful about the choices of
branches when performing computations using the procedure of
\cite{Gopakumars}. However, this subtle issue does not affect the
fact that the prescription of \cite{Gopakumars} always gives
modular-invariant answers.

\subsection{Short edge expansions}
\label{expand}

In this subsection we describe a limit of free field theory
correlation functions which is governed by a saddle point in the Schwinger
parameter space.
The expansion in the position of the saddle point corresponds
to an expansion in the length of one or more small edges in the
critical graph of the corresponding Strebel differential.  As we
will see in the next subsection, this is useful for analyzing
correlators involving diagrams whose contributions are
 localized on subspaces of the
moduli space. These are always diagrams whose dual graph (which is
the critical graph of the Strebel differential) has some vanishing
edges.

For simplicity, consider a four dimensional large $N$ gauge theory,
and consider a correlation function
\begin{equation}<\prod_{i=1}^n \mathcal{O}_i (x_i)>\end{equation}
of gauge-invariant operators made purely of adjoint scalar fields.
In the free gauge theory, each Feynman graph corresponding to such
a correlation function involves $J_{ij}$ contractions of fields
between the $i$'th and the $j$'th operator, so that the answer is
given by
\begin{equation}
\prod_{1 \leq i < j \leq n} |x_i - x_j|^{-2J_{ij}}.
\end{equation}
In the Schwinger parameterization (which generalizes the one
used in \cite{Gopakumars} to position space) this is
rewritten as (up to a constant)
\begin{equation}
\label{Schwingint}
\int \prod_{1 \leq i < j \leq n, J_{ij} > 0}
d\sigma_{ij} \sigma_{ij}^{J_{ij}-1}
e^{-\sigma_{ij} |x_i - x_j|^2},
\end{equation}
and the integration over the $\sigma$'s is identified with the integral
over the decorated moduli space of Riemann surfaces, with
the $\sigma$'s identified with the lengths of the edges of the
critical graph of the Strebel differential. When some of the $J_{ij}$
vanish so that some of the $\sigma$ integrations are not present, this
integral is localized on a subspace of the decorated moduli space, and
it may also be localized on a subspace of the moduli space, as discussed
in \cite{Aharony:2006th}.

Note that a rescaling of all the $\sigma$'s acts on the decorated moduli space just by rescaling the $p_i$'s,
without changing the position on ${\mathcal M}_{g,n}$. Thus, the position on the moduli space is independent of
the overall scale but depends only on the ratios of the $\sigma_{ij}$'s. It is then useful to separate the
integral into the overall length and the ratios. When (say) $J_{12} > 0$, we can do this by defining $s_{ij}
\equiv \sigma_{ij} / \sigma_{12}$. Denoting by $S_1$ the set $\{(i,j) | 1 \leq i < j \leq n, J_{ij} > 0\}$ and
by $S_2$ the set $S_1 - (1,2)$, we can rewrite \eqref{Schwingint} as
\begin{equation}
\label{Schwingintn} \int d \sigma_{12} (\prod_{(i,j) \in S_2}
ds_{ij} s_{ij}^{J_{ij} - 1}) \sigma_{12}^{\Sigma_{(i,j) \in S_1}
(J_{ij} - 1)} e^{-\sigma_{12} (|x_1 - x_2|^2 + \Sigma_{(i,j) \in
S_2} s_{ij} |x_i - x_j|^2)}
\end{equation}
or (up to a constant)
\begin{equation}
\label{Schwingintnn} \int (\prod_{(i,j) \in S_2} ds_{ij}
s_{ij}^{J_{ij} - 1}) \left(|x_1 - x_2|^2 + \sum_{(i,j) \in S_2}
s_{ij} |x_i - x_j|^2 \right)^{ -\Sigma_{(i,j) \in S_1} (J_{ij} -
1) - 1}.
\end{equation}

In general, the integral \eqref{Schwingintnn} is quite complicated
when written as an integral over the moduli space. However, if we look
at the limit when one (or more) of the $J_{ij}$'s becomes very large,
the integral is dominated by a saddle point in the Schwinger parameter
space, and can be evaluated in
the saddle point approximation.  This saddle point maps onto a
point in the decorated moduli space (up to the overall scaling of the
$p_i$), and therefore it is also located at a point in the moduli
space. In the limit of large $J \equiv \sum_{(i,j) \in S_1} J_{ij}$,
and assuming without loss of generality that $J_{12}$ is of order $J$
in this limit,\footnote{At least one of the $J_{ij}$'s must be of order
$J$ in this limit, and we choose the ordering of the points so that it
is $J_{12}$.} this saddle point is located at (at leading order in
$1/J$)
\begin{equation}
s_{ij} = {{(J_{ij}-1) |x_1 - x_2|^2}\over {J_{12} |x_i - x_j|^2}}.
\end{equation}
If all of the $J_{ij}$'s become large then this saddle point will map onto
a point in the middle of the moduli space. However, if one (or more)
of the $J_{ij}$'s remain finite in the large $J$ limit, then the $s_{ij}$ and therefore the
Strebel length $l_{ij}$
of the corresponding edge scales as $1/J$ in the large $J$ limit. Thus,
expanding in $1/J$ leads to an expansion in small edges of the Strebel
differential.

\begin{figure}[htbp]
\begin{center}
\epsfig{file=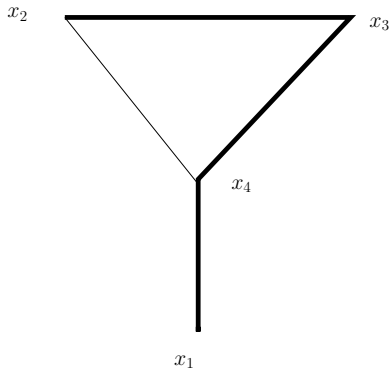, width=2.0in}
\caption{The skeleton graph of the ``Broom'' diagram. The $x_i$ label space time points where operators are
inserted. Bold lines represent many homotopic contractions. On the other hand, the line connecting $x_2$ and
$x_4$ has a small number of contractions. }\label{spade}
\end{center}
\end{figure}

As an example, we can consider the following operators
\begin{gather}
    \mathcal{O}_1(x_1)=\tr(\Phi_1^{J_1}(x_1))\hspace{0.5em},\hspace{0.5em}\mathcal{O}_2(x_2)=\tr(\Phi_2^{J_2}\Phi_3^j(x_2))
\cr
\mathcal{O}_3(x_3)=\tr(\Phi_2^{J_2}\Phi_4^{J_3}(x_3))\hspace{0.5em},\hspace{0.5em}\mathcal{O}_4(x_4)=\tr(\Phi_3^j\Phi_4^{J_3}
\Phi_1^{J_1}(x_4))
\end{gather}
where $\Phi_i$ are some scalar fields in the adjoint of $U(N)$. In this
case the only planar diagram that contributes to
\begin{equation}\label{correlator}<\mathcal{O}_1(x_1)\mathcal{O}_2(x_2)\mathcal{O}_3(x_3)\mathcal{O}_4(x_4)>\end{equation}
has the skeleton drawn in figure \ref{spade}. We can consider the limit where the $J_i$ become large while $j$
remains finite, so that there are many contractions along the bold lines in figure \ref{spade} and a few
contractions on the thin line. The general arguments above imply that in this limit the diagram will get its
dominant contribution from Strebel differentials for which the (dual)
edge corresponding to the line between $x_2$ and
$x_4$ is very small.

The limit where edges of the critical graph of the Strebel
differential vanish is the same as the limit where several zeros
of the Strebel differential come together, leading to a
non-generic critical graph. In the limit where the zeros come
together, this critical graph is the dual of the field theory
diagram which we obtain by removing the edges corresponding to the
finite $J_{ij}$'s from the Feynman graph. Such non-generic Strebel
differentials are often easier to compute. For instance, in the
example described above, the differential for the resulting
``$\Pi$'' diagram was computed in \cite{Aharony:2006th}. We can
then perform a perturbative expansion in the length of the small
edges around these non-generic Strebel differentials, and
interpret the result in terms of the large $J$ limit of the field
theory correlation functions described above.

In this way, we are able to reliably change variables of
integration in some neighborhood of the saddle point of
(\ref{Schwingintnn}). The correlator in the appropriate part of the
Riemann surface (which is the image of a neighborhood of the
saddle point under the Strebel map) is a very good approximation
to the exact result because contributions from other regions in
the field theory integral are substantially smaller.

\subsection{Localized correlators as distributions}\label{loc_dist}

It was shown in \cite{Aharony:2006th} that certain worldsheet correlators, corresponding to non-generic Feynman
diagrams which do not include all possible contractions in their skeleton graph, are localized on
lower-dimension subspaces of the moduli space of Riemann surfaces. This localization may perhaps be attributed
to the fact that the free field theory is a singular limit from the string theory point of view, which is
naively related (at least in the context of string theory on AdS$_5\times S^5$) to the limit in which the
tension of the string vanishes; recall that similar localizations have been encountered also in a different
``tensionless'' limit \cite{Gross:1987ar}. So, this fact by itself is not in contradiction with the existence of
a local worldsheet description. In this subsection we discuss the interpretation of the localized correlation
functions and their implications for the OPE coefficients.

Consider a Feynman diagram giving rise to a localized worldsheet
correlator. This correlator may be written as an integral over a
subspace of the moduli space of the Riemann surface, which is the
subspace spanned by the Strebel differentials whose critical graph has
the topology dual to the Feynman graph. This integral, at least in the
special case considered above of correlation functions of scalar
operators in a four dimensional gauge theory, is positive
definite. This is because the field theory expression for the integral
\eqref{Schwingint} is positive definite, and the mapping of the space
of Schwinger parameters to the decorated moduli space is one-to-one
(leading to a positive definite Jacobian).

Feynman diagrams with a specific topology for their skeleton graph
give localized correlators independently of the number of
contractions; in particular, we can consider the limit where the
number of contractions $J$ is large. In this limit we can consider
adding an additional small number of contractions to the Feynman
diagram such that, as described in the previous subsection, they
will lead to additional small edges in the critical graph of the
Strebel differential, with a length of order $1/J$. Upon adding
the additional edges the correlator is no longer localized, and is
a smooth function on the moduli space for finite $J$ (this is
certainly true if we make all the edges finite, and in many cases
it is true even if we just add one small edge). This function is
still positive, and in the large $J$ limit it becomes localized on
a subspace of the moduli space. Since the localized correlator
arises as a limit of smooth positive functions whose integral is
finite, it is clear that it must be proportional to a
delta-function distribution on the subspace where the correlator
is non-vanishing. While our argument that the correlator should be
thought of as a delta-function distribution is valid only in the
large $J$ limit, it seems likely that the same interpretation of
the localized correlators should hold also for finite values of
$J$.

When we have a correlator localized on a lower-dimension subspace
of the moduli space, in some cases when we take the OPE limit of
two points on the worldsheet approaching each other (say, with two
operators at $\eta$ and at $0$ in the $\eta \to 0$ limit) we find
that the correlator is localized at a specific angle in the
$\eta$-plane, for instance that it is non-zero only for (negative)
real values of $\eta$ (as for the $\Pi$ diagram
\cite{Aharony:2006th}). How is this consistent with having a
smooth OPE ? For simplicity let us consider a four-point function
which is localized on the negative real axis, namely
\begin{equation}\label{localized4pt}<V_1(\eta)V_2(0)V_3(1)V_4(\infty)> =
f(|\eta |)\delta(\theta-\pi)\end{equation}
for some worldsheet operators $V_i$. Here we have written $\eta=|\eta|e^{i\theta}$.
Let us write the OPE expansion of $V_1$ and $V_2$ in the general form
\begin{equation}\label{OPEexpansion}V_1(\eta)V_2(0) \simeq
\sum_{n,\bar{n}}c^{12}_{n\bar{n}}\mathcal{O}_{n,\bar{n}}(0)\eta^{n-2}\bar{\eta}^{\bar{n}-2},
\end{equation}
where the operators $\mathcal{O}_{n,\bar{n}}$ have weight
$(n,\bar{n})$ and we assumed that the $V_i$ are physical operators
of weight $(1,1)$. We know from modular invariance that we can
limit ourselves to integer spins $n-\bar{n}$, though the sums
$n+\bar{n}$ can take any value (and in string duals of conformal
field theories it is believed that they take continuous values
\cite{Aharony:2006th,Brower:2006ea}). Combining
(\ref{localized4pt}), (\ref{OPEexpansion}) and using \be{delt}
2\pi
\delta(\theta-\pi)=\sum_{k=-\infty}^{\infty}(-1)^ke^{ik\theta},
\ee we obtain that for any given value of $n+\bar{n}$ we must have
\begin{equation}\label{OPErelation}
 c^{12}_{n\bar{n}}<\mathcal{O}_{n,\bar{n}}(0)V_3(1)V_4(\infty)>
\sim\text{constant}_{n+\bar{n}}(-1)^{n-\bar{n}}
\end{equation}
where the constant may depend on $n+\bar{n}$ but is independent of
$n-\bar{n}$; namely, for every value of $n+\bar{n}$ there must be
an infinite series of operators appearing in the OPE, with
\begin{equation}\label{nvals}
n={{n+\bar{n}}\over{2}}\ , {{n+\bar{n}}\over{2}} \pm {1\over 2}\ ,
{{n+\bar{n}}\over{2}} \pm 1\ , \cdots,
\end{equation}
which all give rise to the same constant in \eqref{OPErelation}.
A conspiracy of this form (\ref{OPErelation})
is sufficient for reproducing localized correlation functions of the form (\ref{localized4pt}); any operators
appearing in the OPE which are not part of such a conspiracy must not contribute to the localized 4-point
functions.

\section
{The Broom diagram and its limits} \label{Broomlimits}

\begin{figure}\label{broomfig}
\begin{center}
\psfrag{inty}{$\infty$} \psfrag{neta}{$\eta$}
\psfrag{p0}{$p_0$}\psfrag{pinf}{$p_\infty$}\psfrag{peta}{$p_\eta$}
\psfrag{p1}{$p_1$}\psfrag{L1}{$l_1$}\psfrag{L2}{$l_2$}\psfrag{L4}{$l_4$}
\psfrag{C}{$l_3$}
\includegraphics[width=0.5 \linewidth,angle=0]{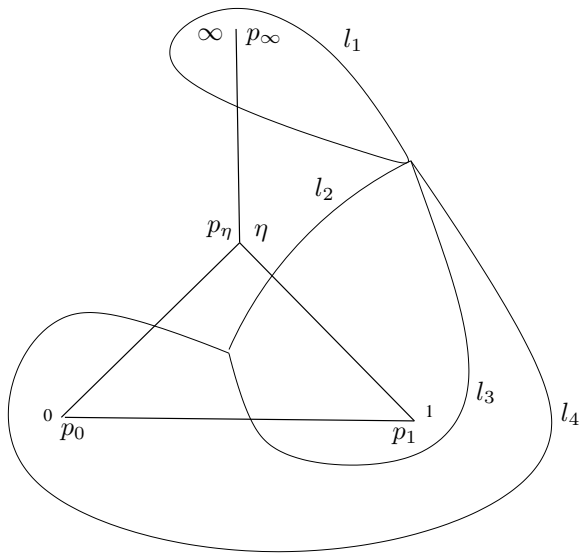}
\caption{The Broom diagram in field theory (the straight lines)
and its dual graph.}
\end{center}
\end{figure}

In this section we study in detail the Broom diagram and the
limits in which it degenerates to the $Y$ and the $\Pi$ diagrams.
In figure \ref{broomfig} we have drawn both the field theory
diagram and its dual diagram which corresponds to the critical
graph of the Strebel differential. The field theory graph is
indicated in bolder lines, while the critical graph is indicated
by curves. The critical graph has two vertices: at one vertex
three lines meet, while at the other five lines meet. Thus, the
Strebel differential corresponding to this graph has one zero of
order 1 and one zero of order 3.

We label the vertices of the Broom diagram to be at $0, 1, \infty,
\eta$ (see figure \ref{broomfig}); these correspond to the poles
of the Strebel differential. The residues at these poles are given
by the positive numbers $p_0, p_1, p_\infty$ and $p_\eta$
respectively. From the figure it is easy to see that the following
relations hold:
\be{curle} l_1 = p_\infty, \qquad l_2 + l_3 = p_1,  \qquad l_4 +
l_3 = p_0,  \qquad l_4 + l_1 + l_2 = p_\eta. \ee
Here $l_1, l_2, l_3, l_4$ denote the Strebel lengths of the edges
of the critical graph as shown in the figure. From the above
equations it is easy to solve  for $l_2, l_3, l_4$ \bea{solleng}
l_2 = \frac{1}{2} \left( p_1 -p_0 + p_\eta -p_\infty\right), \\
l_3 = \frac{1}{2} \left( p_1 + p_0 -p_\eta + p_\infty \right), \\
l_4 = \frac{1}{2} \left( p_\eta - p_\infty - p_1 + p_0 \right).
\eea Thus, all the Strebel lengths are solved in terms of the
residues, which indicates that in order to determine the Strebel
differential of the Broom diagram for given $p_i$ one needs to
solve only algebraic equations, and not the transcendental
equations involving elliptic functions in their full generality.
Note also that we have focussed on a particular Broom diagram, in
which the vertices corresponding to $0$ and $1$ are connected.
There are two other possible Broom diagrams (with a central vertex
at $\eta$), one in which $0$ and $\infty$ are connected and one in
which $1$ and $\infty$ are connected. All these cases are related
by appropriate choices of branch cuts as mentioned in \S
\ref{Genfeat}.

Now let us look at the various limits of the Broom diagram. There
are three limits (for positive $p_i$) and they are obtained by
setting one of the edges $l_2$, $l_3$ or $l_4$ to zero.

\vspace{.5cm} \noindent {\emph{ \underline{$l_2 =0$}}}
\vspace{.5cm}

\begin{figure}
\begin{center}
\psfrag{inty}{$\infty$} \psfrag{neta}{$\eta$}
\psfrag{p0}{$p_0$}\psfrag{pinf}{$p_\infty$}\psfrag{peta}{$p_\eta$}
\psfrag{p1}{$p_1$}\psfrag{L1}{$l_1$}\psfrag{L4}{$l_4$}
\psfrag{L3}{$l_3$}
\includegraphics[width=0.5 \linewidth,angle=0]{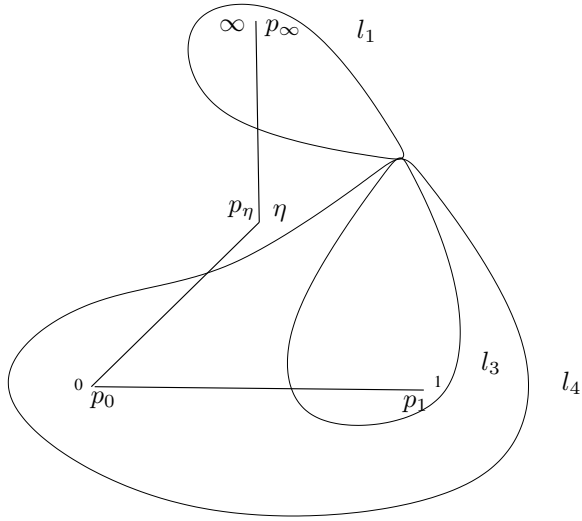}
\caption{The $\Pi_1$ diagram and its dual graph.}\label{C2degenerated}
\end{center}
\end{figure}

As the edge  $l_2$  degenerates, we see the dual edge connecting
$1$ and $\eta$ is removed and one is left with the $\Pi$ diagram
in which the residues satisfy the relation
\be{pi1} p_1 + p_\eta = p_0 + p_\infty. \ee
This $\Pi$ diagram, which we call the $\Pi_1$ diagram, and its
dual graph are shown in figure \ref{C2degenerated}.

\vspace{.5cm} \noindent {\emph{ \underline{$l_3 =0$}}}
\vspace{.5cm}

\begin{figure}
\begin{center}
\psfrag{inty}{$\infty$} \psfrag{neta}{$\eta$}
\psfrag{p0}{$p_0$}\psfrag{pinf}{$p_\infty$}\psfrag{peta}{$p_\eta$}
\psfrag{2}{$p_1$}\psfrag{L1}{$l_1$}\psfrag{L2}{$l_2$}\psfrag{L4}{$l_4$}
\includegraphics[width=0.5 \linewidth,angle=0]{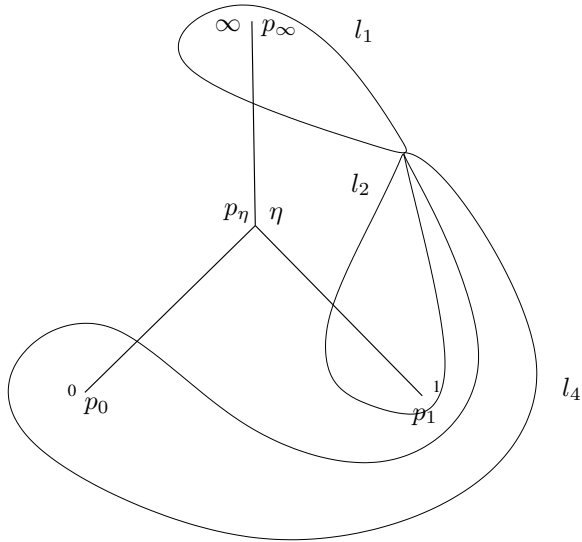}
\caption{The $Y$ diagram and its dual graph.}\label{C3disappears}
\end{center}
\end{figure}

When the edge $l_3$ degenerates, the Broom diagram reduces to the
$Y$ diagram in which the residues satisfy the relation \be{y1} p_0
+ p_1 + p_\infty = p_\eta. \ee Here the dual edge joining $1$ and
$0$ disappears, as depicted in figure \ref{C3disappears}.

\vspace{.5cm} \noindent {\emph{ \underline{$l_4 =0$}}}
\vspace{.5cm}

\begin{figure}
\begin{center}
\psfrag{inty}{$\infty$} \psfrag{neta}{$\eta$}
\psfrag{p0}{$p_0$}\psfrag{pinf}{$p_\infty$}\psfrag{peta}{$p_\eta$}
\psfrag{2}{$p_1$}\psfrag{L1}{$l_1$}\psfrag{L2}{$l_2$}
\psfrag{L3}{$l_3$}
\includegraphics[width=0.5 \linewidth,angle=0]{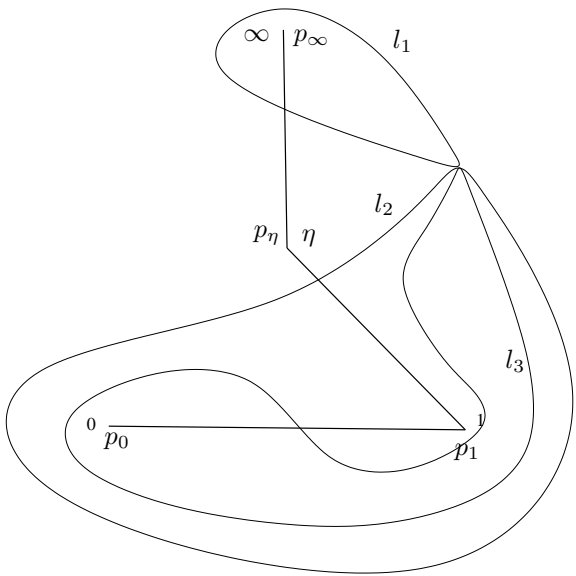}
\caption{The $\Pi_2$ diagram and its dual
graph.}\label{C4disappeared}
\end{center}
\end{figure}

In this limit, the Broom diagram reduces again to a $\Pi$ diagram
which is distinct from the earlier one. Here the residues satisfy
the relation \be{pi2} p_\eta + p_0 = p_\infty + p_1, \ee and the
dual edge joining $\eta$ and $0$ is removed, see figure
\ref{C4disappeared}. We call this diagram $\Pi_2$.

From the above discussion we see that the $\Pi$ diagrams and the
$Y$ diagrams which are obtained as limits of the Broom diagram are
linear subspaces in the space of the $p$'s. In fact, since the
overall scale of the Strebel differential only sets the overall
scale for all lengths, we only need to consider the ratios ${p_i
\over p_\infty}$ $(i=\eta,0,1)$. Then, the $\Pi$ and the $Y$
diagrams correspond to planes in the positive octant of the three
ratios.

Our goal is to compute some Broom diagrams; this should provide us
with new worldsheet correlators which can be studied and lead us
to a better understanding of the worldsheet CFT. Though the
equations are algebraic, it is very hard to obtain closed form
expressions for the cross-ratio $\eta$ as a function of the
lengths $l_i$. As we will see in the next section, at a generic
point in the octant of the $p_i$'s one needs to solve a sixth
order algebraic equation. Thus, closed form expressions can only
be obtained on subspaces where the equations reduce to lower order
ones.

We will therefore mostly use perturbation theory around the limits
mentioned above. There are two different techniques to do this.
One uses straightforward computations on the sphere, while the
other was developed in \cite{David:2006qc} and uses expansions
around special points of elliptic functions. In this section, we
use the more straightforward technique, and in appendix
\ref{pertappendix} we exhibit the other way to do it.\footnote{It
is best to read the appendix after \S \ref{exactsolutions}, where
the relevant schemes are defined.} Of course, the final results
agree precisely. As a further check we will see in the next
section that the perturbation theory agrees with exact results
obtained in some of the exactly solvable subspaces in a common
region of intersection.

\subsection{Perturbing around the $\Pi$ and the $Y$}

In this section we compute explicitly the first non trivial order
of the Broom diagram when expanded around the $\Pi$ and $Y$
diagrams. We begin with the expansion around the $Y$ diagram,
which exhibits many basic features of the perturbation scheme, and
continue to the expansion around the $\Pi$, emphasizing the
implications for the localization of the $\Pi$ found in
\cite{Aharony:2006th}. A detailed field theory analysis using the
world-sheet correlator that we find is postponed to \S
\ref{Fieldtheory}.

\subsubsection{Perturbation around the $Y$ diagram}
\label{pertY}

\begin{figure}\begin{center}
$\begin{array}{c@{\hspace{1in}}c} \epsfig{file=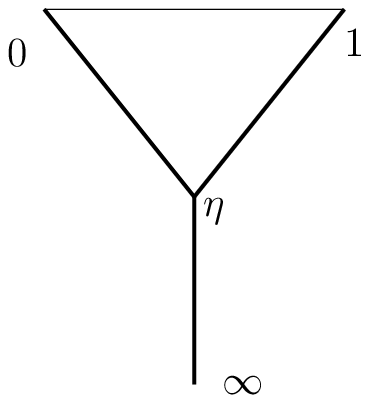, width=1.5in} &
    \epsfig{file=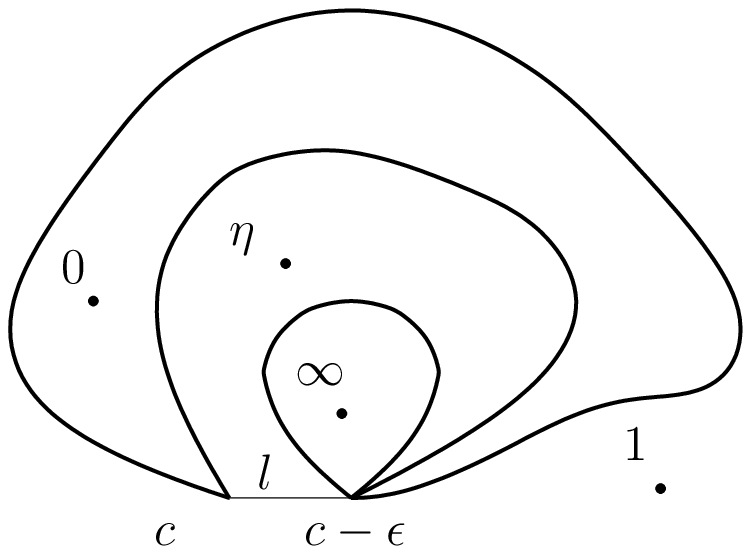, width=2.2in} \\ [0.4cm]
\end{array}$
 \caption{The critical (and the dual) graph for the Broom expanded
 around the $Y$ diagram.
}\label{flag1}\end{center}
\end{figure}

Let us consider the Broom diagram as a perturbation of the $Y$ diagram, and write down its Strebel differential
where we pick the positions of the worldsheet insertions as in figure \ref{flag1}. Our expansion around the $Y$
diagram is an expansion around the subspace where the residues are related by $p_\eta=p_0+p_1+p_\infty$.

We scale all the Strebel lengths relative to $p_\infty$ and define $\gamma_i=p_i/p_{\infty}$. The quadratic
Strebel differential is
\begin{equation}\label{differntial}
qdz\otimes
dz=-p_\infty^2\frac{1}{4\pi^2}\frac{(z-c)(z-c+\e)^3}{z^2(z-1)^2(z-\eta)^2}dz\otimes
dz.
\end{equation}
We consider an expansion around the degenerate differential with
all zeroes coincident, or in other words we perform a Taylor
expansion with small parameter $\e$. As mentioned above, there is
no Strebel condition here, just algebraic constraints which fix
the residues of the poles. The signs of the square roots of the
residues $\gamma_i^2$ of \eqref{differntial} are set by the limit
$\e\rightarrow 0$ which is the $Y$ diagram, for which
\cite{Aharony:2006th} $\gamma_0 = -c^2 / \eta, \gamma_1 = (c-1)^2
/ (\eta-1), \gamma_\eta = (\eta-c)^2 / \eta (\eta-1)$. Expanding
to the first order where the constraint equation is modified we
get

\begin{gather}\label{constraints}
\gamma_0=-\frac{c^2}{\eta}+\frac 32 \frac c{\eta} \e -\frac 38 \frac
1\eta \e^2 -\frac 1{16}\frac{1}{c\eta}\e^3 + O(\e^4),\cr
\gamma_1=\frac{(c-1)^2}{\eta-1}-\frac32
\frac{c-1}{\eta-1}\e+\frac38\frac1{\eta-1}\e^2+\frac
1{16}\frac{1}{(\eta-1)(c-1)}\e^3 + O(\e^4),\cr
\gamma_\eta=\frac{(\eta-c)^2}{\eta(\eta-1)}+\frac32\frac{\eta-c}{\eta(\eta-1)}\e+\frac38\frac{1}{\eta(\eta-1)}\e^2-\frac
1{16}\frac{1}{\eta(\eta-c)(\eta-1)}\e^3 + O(\e^4).
\end{gather}

Let the (rescaled) length of the small edge of the Broom diagram be $l$ (see
figure \ref{flag1}) :
\begin{equation} 2l=\gamma_0+\gamma_1+1-\gamma_\eta. \end{equation}
Using \eqref{constraints} this is equal to
\begin{equation}\label{small
edge} l=\frac 1{32}\frac{\e^3}{c(\eta-c)(c-1)} + O(\e^4).
\end{equation}
Not surprisingly, the edge is non-zero only at the third order in
the perturbation; the fact that the point we expand about has a
four-fold degenerate zero implies that the first two contributions
to the additional edge vanish. Note that in (\ref{small edge}) one
can substitute for $\eta$ and $c$ their zeroth order (in $\e$)
values. The geometric meaning of this equation is very simple : it
is $\frac1{32}\e^3$ multiplied by the residue of the four-fold
zero in the $Y$ diagram. For all other purposes (except
(\ref{small edge})), we may use the equations (\ref{constraints})
to the first non-trivial order in $\e$. This should suffice to
obtain the leading corrections to the amplitude written explicitly
in \cite{David:2006qc}. Some trivial algebra gives the nice result
:
\begin{equation}\label{zero equations}
c=\half \left(1+\gamma_1-\eta(\gamma_0+\gamma_1)+\frac {3\e}{2}\right)+O(\e^3).
\end{equation}
Notice that there is no correction of order $\e^2$. This will play
a major role in the sequel. Substituting this in the first
equation in (\ref{constraints}) we get
\begin{equation}\label{quadratic equation}
    \eta \gamma_0=-\frac{1}{4}(1+\gamma_1-\eta(\gamma_0+\gamma_1))^2+\frac{3\e^2}{16}+O(\e^3).
\end{equation}
It is easy to obtain explicit formulas in the gauge we use, as was
noticed in \cite{David:2006qc}. We can rewrite (\ref{quadratic
equation}) as \begin{equation}\label{convform}
    \eta^2(\gamma_0+\gamma_1)^2-2\eta(\gamma_\eta \gamma_1-\gamma_0)+(1+\gamma_1)^2-3\e^2/4=O(\e^3),
\end{equation}
using the relation among circumferences $\gamma_\eta=\gamma_0+\gamma_1+1+O(\e^3)$. This is a quadratic equation
for $\eta$, with the two solutions corresponding to two possible orientations of the $Y$ diagram before the
additional line is added. We choose the solution of (\ref{quadratic equation}) given by
\begin{equation}\label{right solution}
    \eta=\left(\frac{\sqrt{\gamma_\eta \gamma_1}+
    i\sqrt{\gamma_0}}{\gamma_0+\gamma_1}\right)^2- \frac 3{16}
    \frac{i\e^2}{\sqrt{\gamma_0\gamma_1\gamma_\eta}} + O(\e^3).
\end{equation}
This is the only thing we need, in principle, for the variable
change to the field theory. From here on all the equations we write
will be correct up to $O(\e^3)$. One can invert equation
\eqref{right solution} to simplify the substitution in the field
theory. First, note that
\begin{equation}\label{nright solution}
    \eta-1=\left(\frac{\sqrt{\gamma_1}+
    i\sqrt{\gamma_\eta \gamma_0}}{\gamma_0+\gamma_1}\right)^2- \frac 3{16}
    \frac{i\e^2}{\sqrt{\gamma_0\gamma_1\gamma_\eta}}.
\end{equation}
Simple manipulations lead to the equations
\begin{gather}\label{not conv.}
    (\gamma_0+\gamma_1)|\eta|=(1+\gamma_1)+\frac
    3{32(1+\gamma_1)}\left[-\frac{i\e^2}{\sqrt{\gamma_0\gamma_1\gamma_\eta}}(\sqrt{\gamma_\eta \gamma_1}-
    i\sqrt{\gamma_0})^2+\frac{i\bar{\e}^2}{\sqrt{\gamma_0\gamma_1\gamma_\eta}}(\sqrt{\gamma_\eta \gamma_1}+
    i\sqrt{\gamma_0})^2  \right],\cr
(\gamma_0+\gamma_1)|\eta-1|=(1+\gamma_0)+\frac
    3{32(1+\gamma_0)}\left[-\frac{i\e^2}{\sqrt{\gamma_0\gamma_1\gamma_\eta}}(\sqrt{\gamma_1}-
    i\sqrt{\gamma_\eta \gamma_0})^2+\frac{i\bar{\e}^2}{\sqrt{\gamma_0\gamma_1\gamma_\eta}}(\sqrt{\gamma_1}+
    i\sqrt{\gamma_\eta \gamma_0})^2  \right].
\end{gather}
A nice consistency check is that in the zeroth order, $\e=0$, one
reproduces the equations obtained in \cite{David:2006qc}. It is easy
to solve the equations in the zeroth order (these are linear
equations), and then to iterate and find the first correction. The
zeroth order result is :
\begin{gather}
    \gamma_0=\frac{\abs{\eta-1}-\abs{\eta}+1}{\abs{\eta-1}+\abs{\eta}-1},
    \hspace{2em}
    \gamma_1=\frac{\abs{\eta}-\abs{\eta-1}+1}{\abs{\eta-1}+\abs{\eta}-1},\hspace{2em}
    \gamma_\eta=\frac{\abs{\eta}+\abs{\eta-1}+1}{\abs{\eta-1}+\abs{\eta}-1}.
\end{gather}
To the end of finding the final iterated results as $\gamma=\gamma(\eta,\e)$
the form (\ref{not conv.}) is not very convenient. An equivalent
form of (\ref{not conv.}) (obtained by using the known zero order
relations) is
\begin{gather}\label{da conv}
    (\gamma_0+\gamma_1)|\eta|=(1+\gamma_1)-\frac{3}{16}\frac
    {\abs{\eta-1}+\abs{\eta}-1}{\abs{\eta}}\left( \frac{{\rm Im}( \overline{\e^2} \eta )}{{\rm Im}(\eta)}  \right),\cr
(\gamma_0+\gamma_1)|\eta-1|=(1+\gamma_0)-\frac{3}{16}\frac
    {\abs{\eta-1}+\abs{\eta}-1}{\abs{\eta-1}}\left(\frac{{\rm Im}(\overline{\e^2}(\eta-1))}{{\rm Im}(\eta)}
    \right).
\end{gather}
It is now trivial to solve these equations, and obtain the
corrected form of the circumferences as functions of the sphere
modulus $\eta$ and the separation of the zeroes $\epsilon$. The
first non trivial order of the corrected solution is :
\begin{gather}\gamma_0=\frac{\abs{\eta-1}-\abs{\eta}+1}{\abs{\eta-1}+\abs{\eta}-1}+\frac{3}{16}\frac
    {\abs{\eta}-1}{\abs{\eta-1}}\left(\frac{{\rm Im}(\overline{\e^2}(\eta-1))}{{\rm Im}(\eta)}
    \right)-\frac{3}{16}\frac
    {\abs{\eta-1}}{\abs{\eta}}\left( \frac{{\rm Im}( \overline{\e^2} \eta )}{{\rm Im}(\eta)}
    \right),
    \cr
    \gamma_1=\frac{\abs{\eta}-\abs{\eta-1}+1}{\abs{\eta-1}+\abs{\eta}-1}+\frac{3}{16}\frac
    {\abs{\eta-1}-1}{\abs{\eta}}\left( \frac{{\rm Im}( \overline{\e^2} \eta )}{{\rm Im}(\eta)}  \right)-\frac{3}{16}\frac
    {\abs{\eta}}{\abs{\eta-1}}\left(\frac{{\rm Im}(\overline{\e^2}(\eta-1))}{{\rm Im}(\eta)}
    \right),\cr
    \gamma_\eta=\frac{\abs{\eta}+\abs{\eta-1}+1}{\abs{\eta-1}+\abs{\eta}-1}-\frac{3}{16}\frac
    {1}{\abs{\eta}}\left( \frac{{\rm Im}( \overline{\e^2} \eta)}{{\rm Im}(\eta)}\right)-\frac{3}{16}\frac
    {1}{\abs{\eta-1}}\left(\frac{{\rm Im}(\overline{\e^2}(\eta-1))}{{\rm Im}(\eta)}\right).
\end{gather}

Combining (\ref{small edge}) with (\ref{constraints}) one obtains at
leading order
\begin{equation}\label{length}
    l^2=\frac
1{32^2}\frac{\e^6}{c^2(\eta-c)^2(c-1)^2}=\frac
1{32^2}\frac{-\e^6}{\eta^2(\eta-1)^2}\frac1{\gamma_0\gamma_1\gamma_\eta}.
\end{equation}
The cross-ratio \eqref{right solution} contains $\e^2$, which
means that one has to take a third root of equation
(\ref{length}). This leads to 3 possible solutions which
corresponds to the 3 possible locations of the additional edge.
For any value of $\eta$ and $l$ we have three possible ways to
split the zeros of the differential corresponding to $Y$ diagram
and obtain the Broom, the difference will be in the position of
the extra edge. Fixing the specific diagram we want to compute
uniquely fixes the branch of the cubic root as was explained in \S
\ref{modular invariance section}. We also explained in
\S\ref{modular invariance section} that the naive choice of the
branch is wrong because it changes the worldsheet correlator we
actually compute. This is now obvious from equation (\ref{length})
: a rotation around, say, $\eta=0$ changes the branch upon
computing the cubic root, so it would change the topology of the
diagram. Note that this is consistent only if the additional edge
vanishes somewhere as we do this rotation, and indeed, for
$\eta\in (0,1)$ the circumferences blow up in the zeroth order and
the additional edge, consequently, vanishes.

Finally, to accomplish the variable change we need to integrate
the appropriate field theory integral (\ref{Schwingint}) over the
small edge, $l\equiv s_{01}$. As described in \S\ref{expand}, we
suppose that $\sum_{(i,j)\neq (0,1)} J_{ij}$ is very large
compared to $J_{01}$. Then, this integral is dominated by a saddle
point which is parameterically small $s_{01}\sim J_{01}/\sum J$.
Combining with (\ref{length}) we obtain that
\begin{equation}\label{scaling}
    \frac{J_{01}}{\sum J}\sim \abs{\e}^3,
\end{equation}
which connects the small parameter of the expansion on the Strebel
side to the small parameter in the space-time field theory.

\subsubsection{Perturbation around the $\Pi$ diagram}

\begin{figure}[htbp]
\begin{center}
$\begin{array}{c@{\hspace{1in}}c}
\epsfig{file=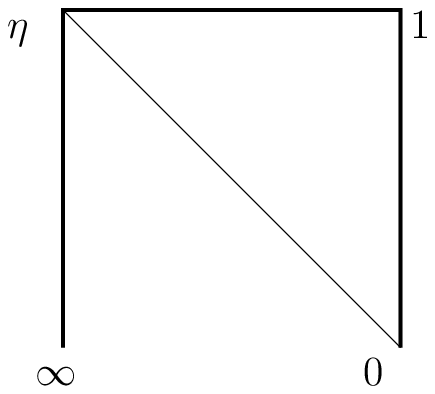, width=1.5in} &
    \epsfig{file=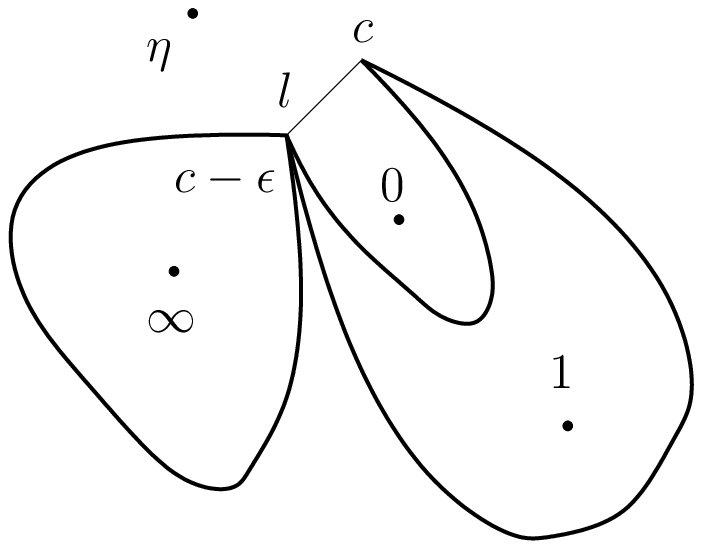, width=2in} \\ [0.4cm]
\end{array}$
 \caption{The skeleton (and the dual graph) of the
field theory graph we consider. The line connecting $0$ and $\eta$
is taken to have few contractions compared to the bold
lines.}\label{flag}
\end{center}
\end{figure}

In this subsection we study the localization of diagrams on the
worldsheet by perturbing around the $\Pi$ diagram.\footnote{More
specifically we work around the $\Pi_2$ diagram.} We start by
discussing the relevant Strebel differential  and then calculate
the field theory expression. Finally, we take a short edge limit,
as discussed in \S \ref{expand}, to get an explicit worldsheet
expression (our short edge is depicted in figure
\ref{flag}).\footnote{There is another possibility of adding an
extra edge to the $\Pi$ diagram by making it into a Square
diagram. However, also for the Square diagram the worldsheet
correlator localizes on the same subspace (see appendix
\ref{squareapp}), so this does not regularize the worldsheet
correlator.}

For the $\Pi$ diagram $p_\infty+p_1-p_0=p_\eta$. We begin with the
differential
\begin{equation}\label{differntial Picase}
qdz\otimes dz=-p_\infty^2\frac1{4\pi^2}
\frac{(z-c)(z-c+\e)^3}{z^2(z-1)^2(z-\eta)^2}dz\otimes dz.
\end{equation}
$\e$ is considered to be small and vanishes for the unperturbed $\Pi$ diagram. As in the expansion around the
$Y$ diagram there are no reality constraints here. Expanding the residues in $\e$ gives
\begin{gather}\label{constraintsPicase}
\gamma_0=\frac{c^2}{\eta}-\frac 32 \frac c\eta \e +\frac 38 \frac 1\eta
\e^2 +\frac 1{16}\frac{1}{c\eta }\e^3 + O(\e^4),\cr
\gamma_1=\frac{(c-1)^2}{\eta-1}-\frac32
\frac{c-1}{\eta-1}\e+\frac38\frac1{\eta-1}\e^2+\frac
1{16}\frac{1}{(\eta-1)(c-1)}\e^3 + O(\e^4),\cr
\gamma_\eta=\frac{(\eta-c)^2}{\eta(\eta-1)}+\frac32\frac{\eta-c}{\eta(\eta-1)}\e+\frac38\frac{1}{\eta(\eta-1)}\e^2-\frac
1{16}\frac{1}{\eta(\eta-c)(\eta-1)}\e^3 + O(\e^4).
\end{gather}
Solving for the zero and substituting back we get
\begin{equation}\label{qudeqPicase}
    \eta^2(\gamma_1-\gamma_0)^2-2\eta(\gamma_\eta \gamma_1+\gamma_0)+(1+\gamma_1)^2-3\e^2/4=O(\e^3),
\end{equation}
which is solved by \begin{equation}\label{solutionPicase}
    \eta=\frac{(\sqrt{\gamma_\eta \gamma_1}\pm\sqrt{\gamma_0})^2}{(\gamma_1-\gamma_0)^2}\pm\frac{3\e^2}{16}\frac1{\sqrt{\gamma_\eta \gamma_1\gamma_0}}
+ O(\e^3).
\end{equation}
It is explicit that in the zeroth order the cross-ratio is purely real. The sign ambiguity appearing in equation
(\ref{solutionPicase}) is fixed by the choice of the  ordering of insertions in the critical graph, figure
\ref{flag} (see \cite{Aharony:2006th} for details). One can show that for our choice of ordering the right sign
choice is $\bold{+}$. Another important equation (all equations from here on are at leading non-trivial order in
$\e$) is
\begin{equation}\label{small edge Picase}
2l=\gamma_\eta+\gamma_0-\gamma_1-1=-\frac 1{16}\frac{\e^3}{c(\eta-c)(c-1)},
\end{equation}
which upon using the relations above can be written in the following
form :
\begin{equation}\label{lengthPicase}
    l^2=\frac
1{32^2}\frac{\e^6}{\eta^2(\eta-1)^2}\frac1{\gamma_0\gamma_1\gamma_\eta}.
\end{equation}
 We can put the zeroth order values of
the $\eta$ modulus in equation (\ref{lengthPicase}). Hence, it
follows that given a specific $l$ there are $6$ solutions for
$\e$, which can be written as :
\begin{equation} \e=e^{\frac{-k\pi i}3
}l^{1/3}\biggl[\frac
1{32^2}\frac{1}{(\eta^{(0)})^2(\eta^{(0)}-1)^2}
\frac1{\gamma_0\gamma_1\gamma_\eta}\biggr]^{-1/6},\hspace{2em}k=0,\cdots,5,
\end{equation}
where $\eta^{(0)}$ stands for the zeroth order solution (which is
a real quantity). Recalling equation (\ref{solutionPicase}) we see
that there are four deformations of the critical graph of the
$\Pi$ diagram for which $\eta$ becomes complex and two for which
it remains real at the leading order.

\begin{figure}[htbp]
\begin{center}
\epsfig{file=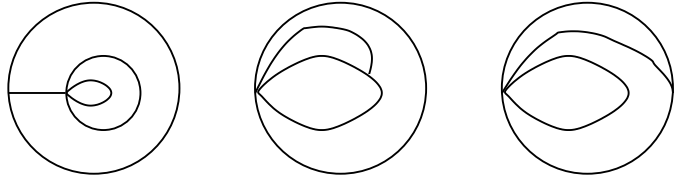, width=3.5in} \caption{3 of the 6
possible blow-ups of the four-fold degenerate zero, the other 3
are similar. The corresponding (dual) field theory graph contains
a self contraction for the figure on the left. The insertions are
at $0$,$1$,$\eta$,$\infty$, and are exactly in this order also for
the faces in the figures. }\label{3 diagrams}
\end{center}
\end{figure}

It is easy to see that there are indeed $6$ possible blow ups of
the $4$-fold degenerate zero in the $\Pi$ diagram which have a
single simple zero and a $3$-fold zero as in \eqref{differntial
Picase}. Two of them are simply adding self contractions to the
$\Pi$ which turn it into a diagram which is set to zero in the
field theory. The other $4$ diagrams correspond to adding a
diagonal line to the $\Pi$. They come in two pairs corresponding
to the mirror image of each other. In general, by imposing $l>0$
in \eqref{small edge Picase} we cut the number of diagrams to $3$,
and then the two complex conjugate solutions of
(\ref{solutionPicase}) correspond to adding a diagonal line (a
Broom diagram, figure \ref{flag}) and the third, real solution,
corresponds to the self contraction\footnote{This may be seen
geometrically, because the original critical curve is symmetric
with respect to complex conjugation. It is possible to add a self
contraction respecting this symmetry (and leaving the Strebel
differential manifestly real), as in figure \ref{3 diagrams},
while it is not possible to do this for the Broom diagram.} of the
vertex mapped to the point $\eta$. These deformations of the
critical graph of the Strebel differential are depicted in figure
\ref{3 diagrams}.

\section{Exactly solvable subspaces of the Broom diagram}\label{exactsolutions}
\subsection
{The Strebel differential for the Broom diagram}

In this section we reanalyze the broom diagram using a different
technique. This uses the uniformizing map for the Strebel
differential onto an auxiliary torus. By taking appropriate
scaling limits of the general case studied in \cite{David:2006qc},
we can specialize to the case of the Broom diagram. We obtain a
set of algebraic equations which turns out in general to be
equivalent to a single sixth order equation. However, we will find
subspaces where the order is lower, where we can obtain closed
form expressions for the cross-ratio. The techniques developed in
this section will be used in appendix \ref{pertappendix} to
rederive the perturbative results of the previous section and to
check some of the exact results obtained here.

As mentioned above, the Strebel differential for the Broom diagram
has both a third order zero  and a simple zero. Therefore, it can
be put in the general form \be{strebsp} \phi(z) dz\otimes dz = -C \frac{ (
za- 1) z^3 }{ ( z-z_0)^2 (z- z_1)^2 ( z-z_2)^2 ( z-z_3)^2 } dz^2.
\ee Here we have chosen the triple zero at the origin and the
simple zero at ${1\over a}$.\footnote{Unlike in the general
discussion and figures of the previous section, the poles are now
at arbitrary positions $z_i$.} The equations which determine the
Strebel differential for a given set of Strebel lengths are
algebraic, as noted earlier. However, we will find it advantageous
to work with an equivalent set of algebraic relations that one
obtains from taking an appropriate limit of the general Strebel
differential for a four-point function.

The most general Strebel differential can be written as
\be{genstreb} \phi(y) dy\otimes dy = - C'\frac{ (y^2 -1) (y^2k^2 -1)}{ (
y-y_0)^2 (y- y_1)^2 ( y-y_2)^2 ( y-y_3)^2 }dy^2. \ee
We substitute the change of variables
\be{cgvar} y= 1 - \epsilon^2 \left( a - \frac{1}{z} \right),
\qquad y_m = 1 - \epsilon^2\left( a - \frac{1}{z_m} \right),
\qquad (m=0,1,2,3)
\ee
and take the limit $\epsilon\rightarrow 0$ with $a$, $y$ and $y_m$ finite. We also scale $C'$ such that
$C'/\epsilon^{10}$ is finite. It easy to see that with this substitution, the general Strebel differential
\eq{genstreb} reduces to that of the Broom diagram given in \eq{strebsp}.

For the general Strebel differential \eq{genstreb}, it is
appropriate to go to the uniformizing variable $u$ for the
elliptic integral
\be{defu} u = \int_1^y \frac{dy}{w}, \qquad w = \sqrt{( 1 -y^2)
(1- y^2k^2)}. \ee
To obtain the required limit we perform the substitution for $y$
given in \eq{cgvar}. We  can then perform the integral by first
expanding the integrand around the origin and then integrating
term by term. This leads to the following asymptotic expansion
\be{expu} u = -\frac{1}{\sqrt{2(k^2-1)}} \left( 2 \epsilon x +
\frac{ 1 - 5k^2}{6 (1-k^2)} (\epsilon  x)^{3} + \frac{ 3 + 2k^2 +
43 k^4}{ 80 ( k^2 -1)} ( \epsilon  x)^{5} + \cdots \right) \ee
where
\be{deftilx}
 x^2 \equiv a - \frac{1}{z}.
\ee
From the above expansion we see that the expansion in $\epsilon$
corresponds to an expansion about $u=0$. To see this in another
way, note that from the definition of $u$ \eq{defu} we have
\be{defzm} y = \frac{\cn (u)}{\dn (u)} \sim 1 + \frac{1}{2} ( k^2
-1) u^2. \ee
Here we have inserted the leading expansions of the functions $\cn
(u)$ and $\dn (u)$ about $u=0$. Inverting the above equation we
see that we obtain the leading expansion in \eq{expu}.

Our goal is to determine the positions of the poles and thus their
cross-ratio $\eta$. The general equations for the positions of the
poles in the $u$ plane are\footnote{See equation (2.9) in
\cite{David:2006qc} and the discussion in \S 2 there. The location
of the pole corresponding to $z_m$ in the $u$-plane is denoted by
$u_m$.}
\be{gencond} \sum_{m=0}^3 r_m \frac{1}{\sn (u_m)} = \sum_{m=0}^3
r_m \sn(u_m) = \sum_{m=0}^3 r_m \frac{\cn(u_m) \dn(u_m) }{\sn(u_m)
} = 0. \ee

The $r_m^2$ are the residues of $\phi(z)$ at $z=z_m$, as in
\cite{David:2006qc}. We can now perform an expansion in $\epsilon$
in these equations to obtain (up to order $\epsilon^6$)
\bea{expcond} & & \sum_{m=0}^3 r_m \left( \sqrt{ 2}{ ( 1-k^2) } {
x_m} + \frac{1 + 3k^3}{ 2 ( k^2-1) } \sqrt{ ( 2 - 2k^2)}
\epsilon^2 x_m^{3} \right) = 0, \cr & &  \sum_{m=0}^3 r_m \left(
\sqrt{\frac{k^2-1}{ 2} } \frac{1}{ x_m} + \frac{1+ 3k^2}{4}\sqrt{
\frac{1}{ 2- 2k^2}} x_m \epsilon^2 + \frac{ (3 + k^2)(5k^2 -1)}{
32 ( k^2 -1)} \sqrt{ \frac{1}{ 2 -
    2k^2}}  x_m^{3} \epsilon ^4 \right) = 0, \cr
& & \sum_{m=0}^3 r_m \left( \sqrt{\frac{k^2-1}{ 2 }}\frac{1}{x_m}
- \frac{ 3 + k^2}{ 4} \sqrt{ \frac{ 1}{2 - 2 k^2}}x_m  \epsilon^2
+ \frac{ 3 k^3 + 50 k^2 -5}{ 16 ( 8 ( k^2 -1))^{3/2}} x_m^{3}
\epsilon ^4 \right) = 0. \eea
The above equations give the following simple conditions, at
leading order, for the poles :
\be{leadcond} \sum_{m=0}^3 r_m { x_m} = \sum_{m=0}^3 r_m
\frac{1}{{\ x_m}} = \sum_{m=0}^3 r_m x_m^{3} = 0. \ee
These equations are algebraic.

The $SL(2,\Cf)$ invariant quantity which characterizes the
location of the poles is their cross-ratio. From the definition of
$x_m$ in \eq{deftilx} and its relation to the poles $y_m$
\eq{cgvar} which occur in the Strebel differential we see that the
cross-ratio is a function of $x_m^2$. Using this freedom one can
suitably change $x_m \rightarrow \pm x_m$ to write the conditions
on the residues as conditions on the perimeters
\be{defperimet} p_0= |r_1| , \quad p_1
=|r_2|, \quad p_\eta= |r_3|, \quad p_\infty =|r_0|. \ee Note that
all the perimeters are greater than zero.  Rewriting the condition
in \eq{leadcond} in terms of the perimeters leads to the following
equations \bea{niccond} \frac{p_0}{x_1} + \frac{p_1}{x_2} +
\frac{p_\eta}{x_3} &=& \frac{p_\infty}{x_0}, \cr p_0 x_1 + p_1 x_2
+ p_\eta x_3 &=& p_\infty x_0, \cr p_0x_1^3 + p_1 x_2^3 + p_\eta
x_3^3 &=& p_\infty x_0^3. \eea

In terms of the ratios
\bea{defgammai}
\gamma_i&=&
\frac{p_0}{p_\infty}, \quad \gamma_1= \frac{p_1}{p_\infty}, \quad
\gamma_\eta= \frac{p_\eta}{p_\infty}, \cr
w_i &=& \frac{x_i}{x_0},
\qquad i =1, 2, 3,
\eea
the conditions \eq{niccond} reduce to
\bea{rdnicecond} \frac{\gamma_0}{w_1} + \frac{\gamma_1}{w_2} +
\frac{\gamma_\eta}{w_3} &=& 1, \cr \gamma_0 w_1 + \gamma_1 w_2 +
\gamma_\eta w_3 &=&1, \cr \gamma_0 w_1^3 + \gamma_1 w_2^3 + \gamma_\eta
w_3^3 &=& 1. \eea Since the perimeters $p_0, p_1, p_\eta, p_\infty$
are positive we have
$\gamma_0, \gamma_1, \gamma_\eta>0$.
Thus we have to look for the solutions to the above
system of equations in the $(\gamma_0, \gamma_1, \gamma_\eta)$
positive octant. These are the equations we will be using in
appendix \ref{pertappendix} to analyze perturbations around some
exact solutions of (\ref{rdnicecond}).

To make the change of variables from the Schwinger parameters to
the cross-ratio we need to express the latter in terms of the
Strebel lengths (which in this case are determined by the
perimeters). The solutions to equations \eq{rdnicecond} are
sufficient to determine the cross-ratio $\eta$ in terms of $p_0,
p_1, p_\eta, p_\infty$
 (or rather the ratios $\gamma_0, \gamma_1, \gamma_\eta$),
\be{fuleta1} \eta = \frac{(w_3^2 - w_2^2)(
w_1^2 -1)}{ (w_1^2 - w_2^2)( w_3^2 -1)}. \ee
Here we have used the
expression for the cross-ratio of the poles $y_i$ and substituted
the change of variables \eq{deftilx}, \eq{cgvar}.

The equations in \eq{rdnicecond} can be reduced to a sixth order
equation in one of the ratios $w_i$. Therefore, it is difficult to
solve them in general. However, they simplify in various limits
which we will discuss next.

\subsection{Global solutions}

As mentioned above, we have to look for solutions to
\eq{rdnicecond} in the positive octant defined by $\gamma_0>0,
\gamma_1>0, \gamma_\eta>0$. In this octant there are three special
planes in which the solutions reduce to that of the $Y$ diagram of
\eq{y1} and the $\Pi$ diagrams of  \eq{pi1} and \eq{pi2}. Since
these planes are important limits of the general solution we will
summarize the results for the cross-ratio in these planes obtained
in \cite{Aharony:2006th, David:2006qc}.

\vspace{.5cm}
\noindent
{\emph{ i. $Y$-plane}}
\vspace{.5cm}

From \eq{y1} we see that the $Y$-plane is specified by the
following linear equation in the octant \be{gamyi} \gamma_0 +
\gamma_1 + \gamma_\eta =1. \ee From
\cite{David:2006qc,Aharony:2006th} , the formula for the
cross-ratio in terms of the $\gamma$'s  reduces to \be{ycross}
\eta_{Y\pm} = \left( \frac{ \sqrt{\gamma_1} \pm i
\sqrt{\gamma_0\gamma_\eta} } {\gamma_0 + \gamma_1} \right)^2. \ee

\vspace{.5cm}
\noindent
{\emph{ii. $\Pi_1$-plane}}
\vspace{.5cm}

We will call the subspace given by \eq{pi1} the $\Pi_1$-plane
\be{gampi1}
 \gamma_1 + \gamma_\eta = 1+ \gamma_0.
 \ee
 The expression
for the cross-ratio is obtained in a way similar to the $Y$-plane
and is given by \be{picross} \ \eta_{\Pi_1\pm} = - \left( \frac{
\sqrt{\gamma_0} \mp \sqrt{\gamma_1\gamma_\eta}
 }{  \gamma_1- \gamma_0} \right)^2.
\ee
 Here we have defined $\eta_{\Pi\pm}$ corresponding to the
$\mp$ branches, note that both branches are real and negative.

\vspace{.5cm}
\noindent
{\emph{ iii. $\Pi_2$-plane}}
\vspace{.5cm}

The $\Pi_2$-plane is the subspace given by \eq{pi2}. In the
$\gamma$-octant, its equation is given by
 \be{gammpi2} \gamma_0 +
\gamma_\eta = 1 + \gamma_1. \ee This plane is related to the
$\Pi_1$-plane by $\gamma_0\leftrightarrow \gamma_1$ exchange. The
equation for the cross-ratio on this plane is given by
\be{pi2cross}
 \eta_{\Pi_2\pm} =  \left( \frac{  \sqrt{\gamma_0}
\pm \sqrt{\gamma_1\gamma_\eta}
 }{  \gamma_1 - \gamma_0} \right)^2.
\ee

An important point to note is that the choice of branches
$\eta_{Y\pm}$ is related to the choice in $\eta_{\Pi_1\pm}$ and
$\eta_{\Pi_2\pm}$. We now focus on two particular limits which
enable us to fix the solution globally, the diagonal line
$\gamma_0=\gamma_1=\gamma_\eta$ and, more generally, the plane
$\gamma_0=\gamma_1$.

\subsubsection{The diagonal line $\gamma_0= \gamma_1=\gamma_\eta=\gamma$}

A nice feature of this line is that it intersects both the
$\Pi$-planes,\footnote{Note, however, that this intersection point
is actually a singular $\Pi$ diagram where one edge is of zero
length.} \eq{pi1}, \eq{pi2}  at $\gamma =1$ and also the
$Y$-plane, \eq{y1} at $\gamma =1/3$. As we will shortly see, on
this plane the equations in \eq{rdnicecond} can be exactly solved
and the cross-ratio can be exactly determined. In addition,
matching with the solutions on the $\Pi$-planes given in
\eq{picross}, \eq{pi2cross} and the $Y$-plane  in \eq{ycross} at
the intersection points will enable us to uniquely fix the branch
of the solution.

We will first evaluate the cross-ratio at the intersection points
with the $\Pi$ and the $Y$ planes. Consider the $\Pi_1$ plane,
and approach the point $\gamma =1$ on the diagonal line along the
$\Pi_1$ plane by taking $\gamma_\eta= 1 + \epsilon, \gamma_1 =
\gamma_0- \epsilon$. Substituting this in \eq{picross} we obtain
to leading order in $\epsilon$
\be{lepicross} \eta_{\Pi+} = - \frac{1}{4\gamma_0} \left( 1
-\gamma_0\right)^2 , \qquad \eta_{\Pi-} = -
4\frac{\gamma_0}{\epsilon^2}. \ee
The values of the cross-ratio given above correspond to the two
different signs in \eq{picross}. We will choose the positive
branch in what follows, a similar analysis can be carried out with
the negative branch. What is important in our analysis is to show
that the perturbation about the $\Pi$ plane leads in general to a
complex cross-ratio; it will be easy to see that this is true in
both of the branches \eq{lepicross}. On the positive branch the
value of the cross-ratio at $\gamma_0=\gamma_1=\gamma_\eta=1$  is
given by \be{fipicross} \eta_{\Pi+} =  - \frac{1}{4\gamma_0}
\left( 1 -\gamma_0\right)^2 =0. \ee Now let us look at the point
$\gamma=1/3$ where the diagonal line intersects the $Y$-plane.
Substituting the values $\gamma_0= \gamma_1 =\gamma_\eta = 1/3$ we
obtain \be{soycross} \eta_{Y\pm} = \exp( \pm i \frac{\pi}{3} ).
 \ee
We again choose the positive branch here since that is what
corresponds to the positive branch in the $\Pi_1$ plane. This
gives the following value for the cross-ratio \be{fiycross}
\eta_{Y+} = \exp(i \frac{\pi}{3} ) = -\omega^2, \ee where $\omega
\equiv \exp(2\pi i /3)$ is the cube root of unity.

The $\Pi_2$ plane is related to the $\Pi_1$ plane by
$\gamma_0\leftrightarrow \gamma_1$. In fact it can be shown from
\eq{picross} and \eq{pi2cross} that $\eta_{\Pi_1\pm}(\gamma_0,
\gamma_1, \gamma_\eta)  = 1- \eta_{\Pi_2\pm}(\gamma_1, \gamma_0,
\gamma_\eta)$. Thus it is sufficient to focus on the $\Pi_1$
plane. Furthermore when $\gamma_0=\gamma_1$ it seems that we have
$\eta_{\Pi_1\pm}$ and $1-\eta_{\Pi_1\pm}$ labelling the same
Riemann surface. This is consistent with the $\Zf_2$ permutation
symmetry of the Broom diagram.

Let us now solve the equations \eq{rdnicecond} at a general point
on the diagonal line. They reduce to the following simple set of
equations
\be{recubieq} \frac{1}{w_1} + \frac{1}{w_2} + \frac{1}{w_3} = w_1
+ w_2 + w_3 = w_1^3 + w_2^3 + w_3^3 = \frac{1}{\gamma}. \ee
One can solve for any of the variables in the above equations to obtain the following cubic equation (we assume
$\gamma\neq 1$) \be{cubiceq} 3 \gamma^2 w_3^3 - 3 \gamma w_3^2 + w_3 -\gamma =0. \ee Here we have chosen to
eliminate $w_1, w_2$ and write the remaining equation for $w_3$. One can solve this cubic equation quite easily:
under the shift $w_3= y + \frac{1}{3\gamma}$, the equation simplifies to \be{simpcubi} y^3 - \frac{1}{3\gamma}
\left( 1 - \frac{1}{9\gamma^2} \right) = 0. \ee This gives the following three solutions  for $w_3$ \be{solx3c}
w_3 = \frac{1}{3\gamma} + \frac{(1, \omega, \omega^2) }{3\gamma} ( 9 \gamma^2 -1)^{1/3}. \ee One can then solve
for the values $w_1, w_2$ from the first two equations of \eq{recubieq}.

From the symmetry of the equations in \eq{recubieq} it is easy to
see that there are $6$ solutions which correspond to the $6$
permutations in the assignments of the $3$ solutions in
\eq{solx3c} to $w_1, w_2, w_3$. From the formula for the
cross-ratio in \eq{fuleta1} one can see that a permutation
involving the exchange $(1\leftrightarrow 2)$ has the effect
$(\eta \leftrightarrow 1-\eta)$, while the exchange
$(1\leftrightarrow 3)$ has the effect $(\eta \leftrightarrow
1/\eta)$.

Since the diagonal line intersects both the $\Pi$ and the $Y$
plane at $\gamma =1$ and $\gamma =1/3$, respectively, we must
choose the assignment such that the cross-ratio reduces to the
values at these points, given in \eq{fipicross} and \eq{fiycross},
respectively. The following assignments satisfy this criterion
uniquely :
\be{thx3sol} w_1= \frac{1}{3\gamma} + \frac{1}{3\gamma}
\nu, \quad w_2 = \frac{1}{3\gamma} + \frac{\omega }{3\gamma} \nu,
\quad w_3 = \frac{1}{3\gamma} + \frac{\omega^2 }{3\gamma}\nu. \ee
Here we have defined, for convenience \be{defnu} \nu \equiv ( 9
\gamma^2 -1)^{1/3} \ee to be the real value of the cube root.
Evaluating the cross-ratio \eq{fuleta1} using the above
assignment, we obtain \be{diacross} \eta_{\rm{dia} } = -
\frac{\omega^2 ( \nu -2)^2 ( \nu +1)} {(\nu-2\omega)^2 (\nu +
\omega)}. \ee It is easy to see that this expression for $\eta$
vanishes at $\nu =2$ ,\footnote{If we had chosen the negative
branch in \eq{lepicross} we would have had to perform the
$w_1\leftrightarrow w_3$ exchange in the assignments of
\eq{diacross}, which would have resulted in $\eta_{\rm{dia}}
\leftrightarrow 1/\eta_{\rm{dia}}$.} or $\gamma =1$, which
corresponds to the point where the diagonal line meets the
$\Pi$-plane. It also correctly reduces to $-\omega^2$ at $\nu=0$
or $\gamma =1/3$, where the diagonal line meets the $Y$-plane.

We easily see from the exact expression \eq{diacross} that, at
least in the direction of this diagonal line, the cross-ratio
becomes complex when we perturb infinitesimally away from the
$\Pi$ plane(s).

\subsubsection{The $\gamma_0=\gamma_1$ plane}

We will now study in a little more detail the behavior of $\eta$
in the plane $\g_0=\g_1\equiv \g$. In this plane we know the exact
value of $\eta$ along the line $\g_{\eta}=1$ (which is the
intersection with the $\Pi$ planes) as well as along the line
$\g_{\eta}=\g$ analyzed above. We will use this exact information
and study the perturbation around the $\Pi$ plane and see that it
is consistent with our general results.

The exact equations \eq{rdnicecond} in this plane take the
simplified form
\be{gampleq} \frac{\g}{w_1} + \frac{\g}{w_2} + \frac{\g_\eta}{w_3}
= \g (w_1 + w_2) + \g_\eta w_3 = \g(w_1^3 + w_2^3) + \g_\eta w_3^3
= 1. \ee
We can eliminate $w_1, w_2$ from these equations and obtain a
quartic equation for $w_3$
\be{gamquart}
\g_\eta(\g_\eta^2-\g^2)w_3^4-\g_\eta^2(\g_\eta^2+3-4\g^2)w_3^3+
3\g_\eta(1+\g_\eta^2-2\g^2)w_3^2
-(3\g_\eta^2+1-4\g^2)w_3+\g_\eta(1-\g^2)=0. \ee
In principle, it is possible to solve this quartic equation
exactly and obtain the cross-ratio explicitly. However, the
general expressions appear to be much too cumbersome, so we will
content ourselves with looking at various limits and, in appendix
\ref{pertappendix}, perturbing around them.

In the limit $\g_\eta=1$ we get the simple quartic equation
\be{pigam} (1-\g^2)(w_3-1)^4=0. \ee For $\g_\eta=\g$, we get the
cubic equation \be{cubic} w_3^3-{1\over \g}w_3^2+{1\over
3\g}w_3-{1\over 3\g}=0, \ee which we had obtained and solved in
the previous subsection. There is a further limit where the
quartic simplifies. This is along the line $\g=1$. We again get a
cubic equation \be{cubic2} \g_\eta w_3^3-\g_\eta^2 w_3^2+3\g_\eta
w_3-3=0. \ee This cubic equation can be mapped to the previous
cubic in terms of the variables  $w_3'={1\over w_3}$,
$\g_\eta'={1\over \g_\eta}$.

\section{Field theory analysis}\label{Fieldtheory}

In this section we return to the field theory and consider the
worldsheet correlator for the Broom diagram, in the limit where it
is close to the $\Pi$ diagram.  As mentioned in \S \ref{Genfeat},
this is obtained by taking a suitable limit of a large number of
contractions. We will, therefore, consider this limit in more
detail.

The field theory integral, written with conductance variables as in \S \ref{expand}, is (up to an overall
numerical factor)
\begin{eqnarray}G(x_0,x_1,x_\infty,x_\eta)&=& \int d\sigma_{\infty\eta}
d\s_{1\eta}d\s_{0\eta}d\s_{01}
   \s_{\infty\eta}^{m_{\infty\eta}}
\s_{1\eta}^{m_{1\eta}}\s_{0\eta}^{m_{0\eta}}\s_{01}^{m_{01}}
\times \\ \nonumber &\;&
\exp\left[-\s_{\infty\eta}(x_\infty-x_\eta)^2-\s_{1\eta}(x_1-x_\eta)^2-\s_{0\eta}(x_0-x_\eta)^2-\s_{01}(x_0-x_1)^2\right]
.
\end{eqnarray}
Notice that the $m$'s are the physical multiplicities $J_{ij}$
defined in \S \ref{expand} minus one, and are non-negative
integers. We integrate over the overall scale to get
\begin{eqnarray}\label{FT integral}
G(x_0,x_1,x_\infty,x_\eta)&=&\int d\sigma_{\infty\eta}
ds_{1\eta}ds_{0\eta}ds_{01}\s_{\infty\eta}^{m_{\infty\eta}+m_{1\eta}+m_{0\eta}+m_{01}+3}
s_{1\eta}^{m_{1\eta}}s_{0\eta}^{m_{0\eta}}s_{01}^{m_{01}} \times
\cr &\;&\exp
\left[-\s_{\infty\eta}\left((x_\infty-x_\eta)^2+s_{1\eta}(x_1-x_\eta)^2+s_{0\eta}(x_0-x_\eta)^2+s_{01}(x_0-x_1)^2\right)
\right] \cr &=& \int ds_{1\eta}ds_{0\eta}ds_{01}
s_{1\eta}^{m_{1\eta}}s_{0\eta}^{m_{0\eta}}s_{01}^{m_{01}} \times
\cr &\;& \left((x_\infty-x_\eta)^2+s_{1\eta}(x_1-x_\eta)^2+
s_{0\eta}(x_0-x_\eta)^2+s_{01}(x_0-x_1)^2\right)^{-M},
\end{eqnarray}
where $s_i\equiv \s_i/\s_{\infty\eta}$ and $M \equiv
m_{\infty\eta} + m_{1\eta} + m_{0\eta} + m_{01} + 4$. We assume
that $m_{0\eta}<<m_i$ where $m_i$ stands for all other $m$'s. The
integration over $s_{0\eta}$ can then be performed by a saddle
point approximation. Most importantly, the saddle point is
dominant and the value of $s_{0\eta}$ at the saddle point is
extremely small. Indeed, denoting
\begin{equation}
A=(x_\infty-x_\eta)^2+s_{1\eta}(x_1-x_\eta)^2+s_{01}(x_1-x_0)^2,
\end{equation}
we have
\begin{equation}\label{saddle value}
    s_{0\eta}\mid_{\text{saddle
    point}}=\frac{m_{0\eta}}{M}\frac{A}{(x_\eta-x_0)^2}.
\end{equation}
Note that around the saddle point we will have an integration of
the form $\int d(\delta s_{0\eta}) e^{-\half \chi (\delta
s_{0\eta})^2}$ with
\begin{equation}
\chi=\frac{M^2}{m_{0\eta}}\biggl[\frac{(x_\eta-x_0)^2}{A}\biggr]^2=
\frac{m_{0\eta}}{\biggl[s_{0\eta}\mid_{\text{saddle
    point}}\biggr]^2},
\end{equation}
and in order for the saddle point to be dominant $\chi$ should be
very large. This is a reason why the small $m_{0\eta}$ expansion
can be useful to study localization. It remains to evaluate the
integral at this saddle point taking into account the quadratic
fluctuations of the Gau{\ss}ian. One can actually evaluate the
integral exactly using the formula
\be{integ_ex} B(x,y)=\frac{\Gamma(x)\Gamma(y)}{\Gamma(x+y)}=
\int_{0}^{\infty}\frac{t^{x-1}}{(1+t)^{x+y}}dt. \ee
The result of the integral over $s_{0\eta}$ is given by
\begin{eqnarray}\label{ftintegral}
G&=&B(m_{0\eta}+1,M-m_{0\eta}-1) \int ds_{1\eta}ds_{01}
s_{1\eta}^{m_{1\eta}} s_{01}^{m_{01}} \times \cr
&&\left((x_\infty-x_\eta)^2+s_{1\eta}(x_1-x_\eta)^2+s_{01}(x_1-x_0)^2
\right)^{-m_{\infty\eta}-m_{1\eta}-m_{01}-3}
\frac{1}{(x_\eta-x_0)^{2(m_{0\eta}+1)}}.
\end{eqnarray}
This effectively means that we have separated out the propagators corresponding to the $m_{0\eta}+1$
contractions, and written the others as in a $\Pi$ diagram. The non trivial input is still to come, with the
transformation to the $\eta$ plane. One should keep in mind that the above form of the integral would be
completely useless in the case where we don't have a dominant saddle, because one cannot generally make the
change of variables in the last equation ($\eta=\eta(s_{1\eta},s_{0\eta},s_{01})$ and $s_{0\eta}$ is generic).

We can now use our results on the Strebel problem from the previous sections to explicitly calculate the
relevant worldsheet correlator, identifying the (normalized) lengths in the two formalisms, $l \sim s_{0\eta}$,
$\gamma_0 \sim s_{01}$, $\gamma_1-\gamma_0 \sim s_{1\eta}$ (in the notation of \S \ref{Broomlimits}). First, let
us summarize the important relations we have established so far : \bea{eqs} &&\e^2=e^{-\frac{2k\pi
i}3 }l^{2/3}\biggl[32^2{(\eta^{(0)})^2(\eta^{(0)}-1)^2}{\gamma_0\gamma_1\gamma_\eta}\biggr]^{1/3},\nonumber\\
&&l\mid_{\text{saddle
    point}}=\frac{m_{0\eta}}{M}\frac{(x_\infty-x_\eta)^2+(\gamma_1-\gamma_0)(x_1-x_\eta)^2+\gamma_0(x_1-x_0)^2}{(x_\eta-x_0)^2},\\
    &&\eta=\frac{(\sqrt{\gamma_\eta \gamma_1}+\sqrt{\gamma_0})^2}{(\gamma_1-\gamma_0)^2}+\frac{3\e^2}{16}\frac1{\sqrt{\gamma_\eta \gamma_1\gamma_0}},
    \quad \gamma_\eta+\gamma_0-\gamma_1=1+2l.\nonumber
\eea We define  ${\rm Re}{[\eta^{(0)}]}=d^2$, to find at zeroth
order in $l$
\begin{equation} \gamma_\eta=\frac{(d+\sqrt{\gamma_0})^2}{d^2-1},\qquad
\gamma_1=\gamma_0+\frac{(d+\sqrt{\gamma_0})^2}{d^2-1}-1=\frac{(1+d\sqrt{\gamma_0})^2}{d^2-1}.\end{equation}
We see that $ d$ is well defined in the range $1\leq d<\infty$.
 In what follows we limit ourselves to small $d^2-1\equiv \a$, since this
 is relevant for the OPE limit $\eta \to 1$ and it will give compact expressions.
However, an analysis for general values of $\a$ is also possible.

The above relations assume smallness of $\e$ and $l$ and thus we can not trust them for all values of
the parameters.
  Smallness of $l$ implies
\begin{equation}l\sim
\frac{m_{0\eta}}{M}\biggl(\frac{x_1-x_\eta}{x_0-x_\eta}\biggr)^2\frac{(1+\sqrt{\gamma_0})^2}{d^2-1}\ll
\text{min}(1,\gamma_0),\end{equation} which constraints $\gamma_0$
(defining $n\equiv m_{0\eta}/M$) to
\begin{equation}\frac{n}{\a}\biggl(\frac{x_1-x_\eta}{x_0-x_\eta}\biggr)^{2}\ll \gamma_0\ll
\frac{\a}{n}\biggl(\frac{x_1-x_\eta}{x_0-x_\eta}\biggr)^{-2}.\end{equation}
For generic, finite  $x_i$ and $\gamma_0$, we see that we can not
get as close as we like to $d=1$; we are bounded by a value of
order $m_{0\eta}/M$. The smallness of $l$ automatically implies
the smallness of $\e$.
 Thus, we conclude that for generic $x_i$ the calculations can be
 trusted for \bea{bounds} n\ll\a\ll 1,\qquad \frac{n}{\a}\ll \gamma_0\ll
 \frac{\a}{n}.\eea

To actually calculate the worldsheet correlator we have to
establish the dictionary between the Schwinger parameters
appearing on the field theory side and the cross-ratio appearing
in the string correlator. From (\ref{eqs}) we find \bea{dict1}
{\rm
Re}{[\eta]}&=&d^2-\frac{3}{2^{5/3}}\biggl(\frac{x_1-x_\eta}{x_0-x_\eta}\biggr)^{4/3}
\frac{(1+\sqrt{\gamma_0})^{2/3}}{\gamma_0^{1/6}}n^{2/3}\a^{1/3},\nonumber\\
{\rm Im}{[\eta]}&=&\pm\frac{3^{3/2}}{2^{5/3}}\biggl(\frac{x_1-x_\eta}{x_0-x_\eta}\biggr)^{4/3}
\frac{(1+\sqrt{\gamma_0})^{2/3}}{\gamma_0^{1/6}}n^{2/3}\a^{1/3}, \eea and we get that (define $\b^{1/3}\equiv
\frac{3^{3/2}}{2^{5/3}}\biggl(\frac{x_1-x_\eta}{x_0-x_\eta}\biggr)^{4/3}$) \be{boundsIm} 2^{2/3}\b^{1/3}
n^{2/3}\a^{1/3}<\i\ll 1 \ee in the trusted regions. From (\ref{dict1}) we can easily write down the dictionary
(choosing the plus sign for ${\rm Im}{[\eta]}$) : \bea{aa} &&\a=\r+\frac{1}{\sqrt{3}}\i-1, \eea as well as the
solution for $\gamma_0$ \bea{p0} \sqrt{\gamma_0^{\pm}}=\frac{\i^3}{2\b\a n^2}-1\pm\sqrt {\frac{\i^3}{2\b\a
n^2}(\frac{\i^3}{2\b\a n^2}-2)}. \eea The plus sign is for $\gamma_0>1$ and the minus sign is for $\gamma_0<1$.
The Jacobian of the transformation $\{\r,\i\}\to\{\gamma_0,d\}$    is \bea{J1} |{\mathcal{J}}|=\mid 2d\frac{\d
\i}{\d \gamma_0}\mid=\b n^{2/3}\a^{1/3}\frac1{3}
\frac{(\sqrt{\gamma_0^{\pm}}+1)^{-1/3}(\sqrt{\gamma_0^{\pm}}-1)}{(\gamma_0^{\pm})^{7/6}},\eea where we have to
sum over both branches for $\gamma_0$. The Jacobian of $\{\gamma_0,\gamma_1\}\to\{\gamma_0,d\}$ is, in the
vicinity of $d=1$, \bea{J2} 2\frac{(1+\sqrt{\gamma_0^{\pm}})^2}{\a^2}.\eea From here we easily obtain that to
leading order the total Jacobian is
\bea{J3}
6\frac{(1+\sqrt{\gamma_0^{\pm}})^{7/3}(\gamma_0^{\pm})^{7/6}}{\b
n^{2/3}\a^{7/3}(\sqrt{\gamma_0^{\pm}}-1)}.\eea

Finally, we can recast  the field theory integral as a worldsheet
expression. The field theory integral \eqref{ftintegral} is
proportional to
\be{simpInt} \int d{\i}
d{\r}\frac{\left(\sqrt{\gamma_0^{+}}\right)^{1/2}}{\sqrt{\gamma_0^{+}}-1}
(\sqrt{\gamma_0^{+}}^{m_{01}-m_{\infty\eta}}-\sqrt{\gamma_0^{+}}^{m_{\infty\eta}-m_{01}})
\a^{\frac{2}{3}+m_{\infty\eta}+m_{01}}
\left(\frac{(1+\sqrt{\gamma_0^+})^{2}}{\sqrt{\gamma_0^+}}\right)^{-\frac{11}{6}-m_{\infty\eta}-m_{01}}.
\ee
Here we have summed over both branches of the Jacobian. One can
use the dictionary (\ref{aa}),(\ref{p0}) to write the integrand
directly in terms of the cross-ratio $\eta$.

\begin{figure}[htbp]
\begin{center}
\epsfig{file=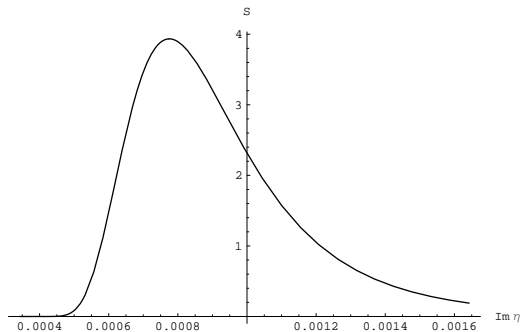, width=2.8in} \caption{A typical
behavior of the integrand. Depicted here is the integrand as a
function of $\i$ for some given $\r$. The vertical axis is
proportional to the amplitude (denoted by $S$). As $m_{0\eta}\to
0$, the integrand approaches $\delta(\i)$.}\label{ImExt}
\end{center}
\end{figure}

We expect that the above integrand has an extremum at the point
corresponding to the unique saddle point of the field theory
integral. We indeed find this maximum, which is a consistency
check on the computation. The typical behavior of the integrand is
shown in figure \ref{ImExt}.

This provides a concrete example of our general discussion in \S
\ref{expand}; we see that in the Broom diagram the correlation
function is a smooth function on the worldsheet, but that in the
limit in which this diagram goes over to the $\Pi$ diagram the
correlator becomes proportional to a delta-function on the real
line.

\section{Summary and discussion}\label{summary and discussions}

In this paper we analyzed in detail a specific four-point function
diagram, the Broom diagram, in order to illustrate several general
features of the translation of \cite{Gopakumars} from Feynman
diagrams to the worldsheet. We showed that this translation is
modular invariant, but that the modular invariance is sometimes
realized non-trivially (requiring following specific branches of
fractional powers). We showed that diagrams which localize on
lower-dimensional subspaces of the moduli space can be thought of
as delta-function distributions, and that such a localization can
be consistent with the worldsheet OPE.

While we have gained some understanding of the localization that
certain worldsheet correlators exhibit, we would also like to be
able to extract some broad patterns related to this behavior. For
instance, we might ask the following general questions:
\begin{itemize}
\item Which free field theory diagrams exhibit localization on a
subspace of the moduli space ?
\item What subspace of the moduli
space do the corresponding Strebel differentials  localize on ?
\end{itemize}
The idea here would be to obtain an answer to the first question
based purely on the graph topology rather than through the kind of
explicit computations that we have carried out. Obviously, a
sufficient condition for localization (discussed in
\cite{Aharony:2006th}) is for a diagram to have less edges than
one plus the (real) dimension of the moduli space, but this is
certainly not a necessary condition. An answer to the second
question would also be important in studying the properties of the
conjectured worldsheet theory.

So far we do not have any general answers to these questions, but
we can make an observation based on the analysis of various sphere
4-point function diagrams in \cite{Aharony:2006th} and in this
paper (including appendix \ref{squareapp}). In these examples two
types of localization occur : localization to a
one-real-dimensional subspace of the two-real-dimensional moduli
space, or localization to a two dimensional region (in the Whale
diagram discussed in appendix \ref{squareapp}). In all cases, the
localization is such that one of the three possible OPE limits of
the diagram (which are $\eta \to 0$, $\eta \to 1$ and $\eta \to
\infty$) does not appear (namely, the region covered by the
Strebel map comes close to two of the fixed vertices but not the
third one). Recall that in a 4-point function one cannot
distinguish between the OPE limit of two points coming together
and that of the other two points coming together (the two limits
are related by a modular transformation). In all cases, we find
that the worldsheet OPE limit that is not present corresponds to
bringing two points together such that both pairs of the
corresponding space-time operators have no contractions between
them in the Feynman diagram that is being computed. For example,
in the $\Pi_1$ diagram of figure \ref{C2degenerated}, the covered
region includes $\eta \to 0$ and $\eta \to \infty$ but not $\eta
\to 1$, and there are indeed no contractions between the operators
at $\eta$ and at $1$, as well as between the operators at $0$ and
at $\infty$.

We can give the following heuristic explanation for this
observation. In general, the space-time OPE of single-trace vertex
operators in large $N$ gauge theories contains two types of terms
(which contribute at leading order in large $N$ to correlation
functions) : single-trace operators and double-trace operators. In
free large $N$ theories, the single-trace operators arise just by
contractions, and they do not contribute to diagrams where there
is no contraction between the two operators in the OPE. Now, in
general, as discussed in \cite{Aharony:2006th}, there is no clear
relation between the space-time OPE and the worldsheet OPE in the
string theory corresponding to a large $N$ gauge theory. However,
when we have an OPE between two operators in some $n$-point
function, the single-trace terms in the space-time OPE are related
to $(n-1)$-point functions, which are in turn related to
$(n-1)$-point functions on the worldsheet, so it is natural to
expect that the contributions related to these terms in the
space-time OPE will arise from the OPE limit also on the
worldsheet (which also gives $(n-1)$-point functions). It is hard
to directly relate the two OPEs because the operators appearing in
the space-time OPE are non-normalizable (from the point of view of
the string theory living in a higher dimensional space), and such
non-normalizable single-trace operators do not appear in the
worldsheet OPE;\footnote{The question of how the worldsheet OPE is
related to the space-time OPE in the AdS/CFT correspondence can be
explicitly studied in the AdS$_3$/CFT$_2$ case (for some of the
relevant works see \cite{ads3cft2}).} however, it seems likely
that some relation of this type should exist. The observation in
the previous paragraph is consistent with this point of view,
since the only cases where a worldsheet OPE does not exist is when
it is not required to exist from the point of view of the
space-time OPE. However, the observation is stronger, since one
might think that a worldsheet OPE could exist even when it is not
required to exist, but (in the examples that we analyzed) this
does not happen. This seems to suggest that the worldsheet OPE
always leads to non-zero contributions related to the single-trace
operators in the space-time OPE, so that the appearance of the two
is precisely correlated.

It would be interesting to check whether the observation described
above can be generalized also to higher $n$-point functions, and
whether all localizations may be explained by arguments like the
one presented in the previous paragraph. It seems that more
elaborate arguments are needed to explain localizations of the
type found in two-point functions on the torus in
\cite{Aharony:2006th}. It would also be interesting to understand
why in some cases the diagram localizes to a lower dimensional
subspace of the moduli space, while in other localized cases it
has a non-zero measure on the moduli space. One difference between
these cases is that the localization to the lower dimensional
subspace implies strong constraints on the worldsheet OPE limit
(for pairs of points which do have an OPE limit), as discussed in
\S \ref{loc_dist}, while no such constraints arise when the
localized subspace has non-zero measure. Note that when we go away
from the free field theory limit we expect all of these
localization properties to disappear, since at some order in the
perturbation expansion in our coupling constant there should
appear some diagram which covers the whole moduli space; this is
consistent with the argument above, since in a non-free theory
single-trace operators in the OPE can always contribute to any
correlation function, beginning at some order in the perturbation
expansion.

\acknowledgments We would like to thank O. Bergman, M. Berkooz, D. Gross, N.
Itzhaki, A. Neitzke, I. Yaakov and A. Yarom for useful
discussions. O.A and Z.K would like to thank the Institute for
Advanced Study for its hospitality during the course of this
project. The work of O.A. and Z.K. was supported in part by the
Israel-U.S. Binational Science Foundation, by the Israel Science
Foundation (grant number 1399/04), by the European network
HPRN-CT-2000-00122, by a grant from the G.I.F., the German-Israeli
Foundation for Scientific Research and Development, and by a grant
of DIP (H.52). J.R.D. and R.G. are very grateful to the generous
support extended to their work by the people of India. J.R.D. also
thanks the string theory group at the Weizmann Institute for kind
hospitality. We also thank the organizers of the Indo-Israeli
meeting at Ein Boqeq where this work was initiated. The work of
S.S.R. is supported in part by the Israel Science Foundation under
grant no. 568/05.

\appendix

\section{Perturbation expansion and the matching with exact answers}\label{pertappendix}

Our goal here is to find the particular branch of the perturbation
expansion around the $\Pi$-plane which is consistent with the
global solutions discussed in \S \ref{exactsolutions}. We will
carry out the general expansion in appendix
\ref{appendixperturbations}. However, as a simpler exercise we
will first perturb in the $\gamma_0=\gamma_1=\gamma$ plane. This
will illustrate how matching with the exact solution along the
diagonal line fixes the branch.

\subsection{Perturbing around the line $\gamma_\eta =1$}

With the data of the exact solution in these limits we can study
the perturbation around the line $\g_\eta=1$. More precisely let
us consider the strip $\g_\eta=1+\ep$ and fixed $\g$. To study the
perturbation, we need to make a double expansion of \eq{gamquart}
in $\ep$ and $\delta w_3=w_3-1$.  The resulting equation is
\be{quartpert} (\g^2-1)\delta w_3^4+8\g^2 \ep \delta w_3+12\g^2\ep
\delta w_3^2+4\g^2\ep^2 +12\g^2\ep^2\delta w_3 + \ldots=0. \ee
If we take $\delta w_3 \propto \ep^\beta$, we see that the first
few terms scale as $\ep^{4\beta}, \ep^{\beta+1}, \ep^{2\beta+1},
\ep^2, \ep^{2+\beta}$ respectively. It is easy to verify that we
can have two consistent branches for the perturbation. In one
branch $\ep^{\beta+1} \propto \ep^2$, implying $\beta=1$. The
other is where $\ep^{4\beta}\propto \ep^{\beta+1}$, which implies
that $\beta={1\over 3}$.

In the first expansion where $\delta w_3\propto \ep$, we can show from
\eq{fuleta1} that
$\eta \propto {1\over \ep^2}$.  This is not the branch we need.

In the second expansion where $\delta w_3\propto \ep^{1\over 3}$,
we find upon solving \eq{quartpert} to leading order that
\be{delw3pert} \delta w_3^3=-{8\g^2\over (1-\g^2)}\ep \equiv
-8\mu^3\ep. \ee Therefore, $w_1w_2=\frac{1-\g_\eta
w_3}{w_3-\g_\eta}w_3=-1+2\mu\ep^{1\over 3}$. Here $\mu$ has to be
chosen to be one of the appropriate cube roots defined by
\eq{delw3pert}, as will be fixed later. Together with the equation
for $w_1+w_2$, this gives
\be{deltw1w2} \delta w_1= \mu\ep^{1\over 3}\frac{1-\g}{\g}, \qquad
\delta w_2= \mu\ep^{1\over 3}\frac{1+\g}{\g}. \ee
This gives a cross-ratio (to leading order) \be{etapi}
\eta=-{1\over 4\g}(1-\g)^2, \ee which agrees with the answer
obtained for the $\Pi$ plane. This is, therefore, the correct
branch of the perturbation expansion.

However, we see from the expression \eq{delw3pert} for $\mu$ that
this perturbation expansion is valid away from $\g=1$. it will
break down in some vicinity of $\g=1$.  More precisely, if we take
$\g=1+\delta$ for small $\delta$, then we see from \eq{delw3pert}
that the small parameter in the perturbation expansion is not
$\ep$ but rather ${\ep \over \delta}\equiv {1\over b}$. We can
thus approach $\g=1$ as long as we choose $\ep$ so that $b \gg 1$.

Actually, we can independently do a perturbation expansion for
arbitrary values of $b$. In other words consider some small
neighborhood of $\g=\g_\eta=1$. We take $\delta=b\ep$ and do a
double expansion of \eq{gamquart} in $\delta$ and $\ep$ keeping
the first few terms. We get \be{aexp}
2\ep(w_3-1)[b(w_3-1)^3-w_3(w_3^2+3)]=0. \ee Since $w_3$ is not
identically one everywhere in this region of expansion, we have
the cubic equation for $w_3$ \be{w3cub} \alpha
w_3^3+3w_3^2+3\alpha w_3+1=0, \ee where $\alpha \equiv -{b-1\over
b}$. The roots of this cubic equation are given by \be{w3roots}
w_3= \{ -{1\over \alpha}(q+r+1), -{1\over
\alpha}(q\omega+r\omega^2+1), -{1\over \alpha}(q\omega^2
+r\omega+1) \}, \ee where
\be{defqr} q \equiv (1-\alpha^2)^{1\over 3}(1+\alpha)^{1\over 3},
\qquad r \equiv (1-\alpha^2)^{1\over 3}(1-\alpha)^{1\over 3}. \ee
In the $b\rightarrow \infty$ limit,
 we see that $q\sim 2^{1\over 3}/ b^{2\over 3}$ and
$r\sim 2^{2\over 3}/ b^{1\over 3}$,
 so that to leading order in ${1/ b}$
 \be{w3a}
w_3= \{ 1+{2^{2\over 3}\over b^{1\over 3}},
1+{2^{2\over 3}\over b^{1\over 3}}\omega^2,
1+{2^{2\over 3}\over b^{1\over 3}}\omega \}.
\ee
This exactly matches with the
result of the earlier perturbation expansion \eq{delw3pert}.

To decide which is the appropriate cube root to pick we go back to
the exact answers obtained on the line $\g_\eta=\g$.  By
specializing to the case $b=1$, we restrict ourselves to this line
in the vicinity of $\g_\eta=\g=1$. In this case the possible
values of $w_3$ from \eq{w3roots} are given by \be{w3a1} w_3=\{
\infty , -{i\over \sqrt{3}}, {i\over \sqrt{3}} \}. \ee But from
the exact answer we know that the global properties demand that we
choose the middle root $w_3=-{i\over \sqrt{3}}$ (as in
\eq{thx3sol}). Since we expect to continuously interpolate between
$b=1$ and large $b$, we must continue to pick the middle root in
the perturbation expansion about the $\Pi$-plane in \eq{w3a} as
well. As we will see in the next subsection this choice leads to
the cross-ratio near the $\Pi$-plane being generically complex.

\subsection{Perturbation expansion about the $Y$ and the $\Pi$ diagrams}\label{appendixperturbations}

We now consider the perturbation expansion about a generic point
in either the $\Pi$ or the $Y$-planes. The exact global solutions
discussed in the previous sections will fix the precise branch of
the solution. This  will enable us to show that  the perturbation
expansion  of the cross-ratio about the $\Pi$-plane is generically
complex though the cross-ratio on the $\Pi$-plane is real.

We first obtain a perturbation expansion either about the $Y$ or
the $\Pi$ diagram by looking at a further limit of the equations
in \eq{leadcond}. From \eq{strebsp} we see that in the limit
$a\rightarrow\infty$ we obtain the appropriate Strebel
differential either for the $Y$ or the $\Pi$ diagram. Expanding
the equations \eq{leadcond}, using  \eq{deftilx} in the
$a\rightarrow\infty$ limit we obtain \bea{ainftycon} \sum_{m=0}^3
r_m \left( 1 - \frac{1}{2z_m a} -\frac{1}{8 (z_ma)^2} -
\frac{1}{16} \frac{1}{(z_ma)^3} + \cdots \right) &=& 0, \cr
\sum_{m=0}^3 r_m \left( 1 + \frac{1}{2z_m a} +\frac{3}{8 (z_ma)^2}
+ \frac{5}{16} \frac{1}{(z_ma)^3} + \cdots \right) &=& 0, \cr
\sum_{m=0}^3 r_m \left( 1 - \frac{3}{2z_m a} +\frac{3}{8 (z_ma)^2}
+ \frac{1}{16} \frac{1}{(z_ma)^3} + \cdots \right) &=& 0. \eea
Taking linear combinations of these equations further we obtain
the following equivalent set
\be{equainc} \sum_{m=0}^3 r_m \left( 1 + \frac{1}{16} \tilde
\epsilon^3 \tilde z_m^3 \right) + O(\tilde\epsilon^4) =
\sum_{m=0}^3 r_m \left( \tilde z_m  + \frac{1}{8} \tilde
\epsilon^2 \tilde z_m^3 \right) + O(\tilde\epsilon^3) =
\sum_{m=0}^3 r_m \left( \tilde z_m^2 + \frac{1}{2} \tilde \epsilon
\tilde z_m^3 \right) + O(\tilde\epsilon^2) = 0, \ee
where $\tilde \epsilon = 1/a$ and $\tilde z_m = 1/z_m$. From these
equations we see that the leading order equations reduce to those
found in \cite{David:2006qc} (see equation (3.4)). To solve for
the corrected cross-ratio to order $\tilde\epsilon$  we  first
remove the translational mode $\tilde z_0$ by defining $\tilde z_1
= \tilde v_1 + \tilde z_0$, $\tilde z_2 = \tilde v_2 + \tilde
z_0$, $\tilde z_3 = \tilde v_3 + \tilde z_0$. In terms of these
variables the sum of the residues in the first equation of
\eq{equainc} reduces to \be{newsumr} \sum_{m=0}^3 r_m = -
\frac{\tilde\epsilon^3}{16} ( r_1 \tilde v_1^3 + r_2 \tilde v_2^3
+ r_3 \tilde v_3^3). \ee Here we have kept terms only to
$O(\tilde\epsilon^3)$. From this equation we see that the
condition on the sum of the residues is violated only at
$O(\tilde\epsilon^3)$. Keeping terms to $O(\tilde\epsilon^2)$ in
the
 remaining two equations of \eq{equainc}
we obtain \bea{perteq} r_1 \tilde v_1 + r_2 \tilde v_2 + r_3
\tilde v_3 &=& -\frac{\tilde\epsilon^2}{8} ( r_1 \tilde v_1^3 +
r_2\tilde v_2^3 + r_3 \tilde v_3^3 ), \cr r_1 \tilde v_1^2 + r_1
\tilde v_2^2 + r_3 \tilde v_3^2 &=& -\frac{\tilde\epsilon}{2}( r_1
\tilde v_1^3 + r_2 \tilde v_2^3 + r_3 \tilde v_3^3 ) +
O(\tilde\epsilon^2). \eea Note that in the last equation we have
not evaluated the $O(\tilde\epsilon^2)$ term explicitly. We will
see that this term does not contribute to the cross-ratio to
$O(\tilde\epsilon^2)$. Now define \be{defratio} v_1 = \frac{\tilde
v_1}{\tilde v_3}, \qquad v_2 =\frac{\tilde v_2}{\tilde v_3}. \ee
Then the equations \eq{perteq} reduce to \bea{redperteq} r_1 v_1 +
r_2 v_2 + r_3 &=& -\frac{\tilde\epsilon^2}{8} \left( r_1 v_1^3 +
r_2 v_2^3 + r_3\right) \tilde v_3^2 , \cr r_1 v_1^2 + r_2 v_2^2 +
r_3 &=& - \frac{\tilde\epsilon}{2} \left( r_1 v_1^3 + r_2 v_2^3 +
r_3\right) \tilde v_3 + O(\tilde\epsilon^2). \eea It is clear from
these equations that they reduce to the zeroth order equations of
\cite{David:2006qc} (see equation (3.5)), and we need the
information of $\tilde v_3$ at the zeroth order to solve for the
first corrections in $v_1, v_2$. The zeroth order expression for
$\tilde v_3$ is obtained from the deviation of the sum of
perimeters. Writing \eq{newsumr} in terms of $v_1, v_2$ and
$\tilde v_3$ we obtain \be{zw3}
 - 2\sum_{m=0}^3 r_m = \frac{\tilde\epsilon^3}{8} ( r_1 v_1^3 + r_2 v_2^3 + r_3)
 \tilde v_3^3.
\ee Writing $ \sum_m r_m = s \tilde \epsilon^3$ we obtain $\tilde v_3$ to the zeroth order as \be{w3zeroth}
\tilde v_3^{(0)} = - 2 (1, \omega, \omega^2) \left( \frac{ 2 s}{ r_1 v_1^{(0)3} + r_2 v_2^{(0)3} + r_3}
\right)^{1/3}. \ee Here the superscripts in $v_1$ and $v_2$ refer to their zeroth order values. We will fix the
choice of the  cube root later using global considerations. Let us define \bea{defdelt} \delta_1 & =&
\frac{1}{2}
        ( r_1 v_1^3 + r_2 v_2^3 + r_3) \tilde v_3
+ O(\tilde\epsilon), \cr \delta_2 &=&  \frac{1}{8} ( r_1 v_1^3 +
r_2 v_2^3 + r_3) \tilde v_3^2. \eea Then, the two equations of
\eq{redperteq} are \bea{simppereq} r_1 v_1 + r_2 v_2 + r_3 =
-\tilde\epsilon^2 \delta_2 , \cr r_1 v_1^2 + r_2 v_2^2 + r_3 =
-\tilde\epsilon \delta_1. \eea We will show that the correction to
the cross-ratio is a function of $\tilde\epsilon^2 \delta_1$ and
$\tilde \epsilon^2 \delta_2$, thus the $O(\tilde\epsilon)$
correction to $\delta_1$ in \eq{defdelt} is not required at this
order. Eliminating $v_2$ using the first equation of
\eq{simppereq} we obtain the following quadratic equation for
$v_1$, \be{quadw1} r_1(r_1 +r_2) v_1^2 + 2r_1 ( r_3 +
\tilde\epsilon^2\delta_2) v_1 + r_3( r_3 +r_2 + 2 \tilde\epsilon^2
\delta_2) + r_2 \tilde\epsilon\delta_1 =0. \ee Here we have
retained terms to $O(\tilde\epsilon^2)$; note  that $\delta_1$ has
an $O(\tilde\epsilon)$ term which will not be important in the
final result for the  cross-ratio. Solving for $v_1$ to
$O(\tilde\epsilon)$  we  obtain \be{finsolw1} v_1 = \frac{ -r_1(
r_3 + \tilde\epsilon^2 \delta_2) \pm \sqrt{D} }{ r_1( r_1 + r_2)
}, \ee where \be{defD} D= r_1 r_2 r_3 r_0 - r_1 r_2 (r_1+ r_2)
\tilde\epsilon\delta_1 - 2r_1 r_2 r_3 \tilde\epsilon^2 \delta_2.
\ee From the first equation of \eq{simppereq} we obtain
\be{finsolw2} v_2 = -\frac{ r_2 ( r_3 + \tilde\epsilon^2 \delta_2)
\pm \sqrt{D} }{ r_2( r_1 + r_2)}. \ee The cross-ratio is obtained
from \be{defcrossr} \eta = v_1 \frac{ 1 - v_2}{ v_1 -v_2}. \ee
Substituting the values of $v_1$ and $v_2$ from \eq{finsolw1} and
\eq{finsolw2} we obtain
\be{crossans} \eta_\pm= \pm \frac{1}{( r_1 + r_2)^2
\sqrt{D} } \left( -r_1( r_3 + \tilde\epsilon^2 \delta_2) \pm
\sqrt{D} \right) \left( r_2( -r_0 + \tilde\epsilon^2 \delta_2) \pm
\sqrt{D} \right). \ee
Now, expanding all terms to $O(\tilde\epsilon^2)$ we obtain the
following expression for the cross-ratio
\be{crossexp} \eta_\pm = \eta_{\pm}^{(0)} \pm
\frac{1}{\sqrt{r_1r_2r_3 r_0}} \tilde\epsilon^2 \delta_2 \left(
r_1 r_2 ( r_0-r_3) \pm (r_2 -r_1) \sqrt{r_1 r_2 r_3 r_0} \right)
\pm \frac{\tilde\epsilon^2 \delta_1^2}{4}
\sqrt{\frac{r_1r_2}{r_3r_0} } \frac{1}{r_3r_0}, \ee
where the zeroth order term for the cross-ratio is given by
\be{zerocross} \eta^{(0)}_{\pm} = \pm \frac{1}{(r_1 + r_2)^2
\sqrt{r_1r_2r_3r_0} } \left( - r_1r_3 \pm \sqrt{r_1r_2r_3 r_0}
\right) \left( - r_2 r_0 \pm \sqrt{r_1r_2 r_3 r_0} \right). \ee
Note that the corrections to the cross-ratio begin at
$O(\tilde\epsilon^2)$, and the $\delta_1$ dependence occurs as
$\tilde\epsilon^2 \delta_1^2$. Therefore, we need only the leading
term in the $\tilde\epsilon$ expansion of $\delta_1$. We now have
to substitute the values of $\delta_1$ and $\delta_2$. This is
simplified by the observation \be{goodrel} r_1 v_1 ^{(0)3} + r_2
v_2^{(0)3} + r_3 = \frac{r_3 r_0}{ r_1  r_2 ( r_1 + r_2)^2} \left(
 r_1r_2 ( r_0 - r_3 ) \pm \sqrt{r_1 r_2 r_3 r_0} ( r_2 -r_1)
\right). \ee Using this relation one can write the term involving
$\delta_2$  in corrections to the cross-ratio in \eq{crossexp} in
terms of $\delta_1^2$. Then the expression for the cross-ratio
simplifies to \be{crossexps} \eta_\pm = \eta_{\pm}^{(0)} \pm
\frac{3}{4} \sqrt{\frac{r_1 r_2}{r_3r_0}} \frac{1}{r_3r_0}
\delta_1^2 \tilde\epsilon^2. \ee
Now, substituting the value of $\delta_1$
from \eq{defdelt} and using the relation \eq{goodrel} we obtain
\be{fincrosexp} \eta_\pm  = \eta_{\pm}^{(0)} \pm \frac{3}{16} ( 1,
\omega^2, \omega ) \frac{2^{8/3} (s\tilde\epsilon^3)^{2/3}}{ (r_1
r_2r_3r_0)^{1/6} (r_1 + r_2)^{8/3} } \left( \sqrt{ r_1r_2} ( r_0
-r_3) \pm \sqrt{r_3r_0} ( r_2 -r_1) \right)^{4/3}. \ee

We now need to choose the value of the cube root of unity in
$\tilde v_3^{(0)}$. We do this by systematically matching the
solutions globally. Since we have chosen the positive branch of
the square root we must choose the branch $\eta_{+}$. We fix the
choice of the cube root by examining the cross-ratio at the point
$\gamma_0=\gamma_1=\gamma_\eta = 1/3+ \tilde\epsilon^3$. This
point lies on the diagonal line, and it is also close to the
$Y$-plane. Evaluating the cross-ratio from \eq{diacross} we obtain
\be{crosdiaexp} \eta_{{\rm dia}} = -\omega^2 + i \frac{3^{13/6}}{
4^{2/3}} \omega \tilde \epsilon^2. \ee One can easily verify that
it is the choice $\omega^2$ among the roots in \eq{fincrosexp} in
which the cross-ratio $\eta_+$ reduces to the above expression.
Thus, the cross-ratio at a generic point near the $Y$-plane is
given by
\be{genycrosp} \eta_{Y+} = \eta_{Y+}^{(0)} + i\frac{3}{16} \omega
\frac{2^{8/3} (s
\tilde\epsilon^3)^{2/3}}{(\gamma_0\gamma_1\gamma_\eta)^{1/6}
(\gamma_0+\gamma_1)^{8/3} } \left( \sqrt{ \gamma_0\gamma_1} ( 1
+\gamma_\eta ) +  i \sqrt{\gamma_\eta} ( \gamma_0 -\gamma_1)
\right)^{4/3}, \ee
where $ s\tilde\epsilon^3 = \gamma_0 + \gamma_1 + \gamma_\eta -1$
and $\eta_{Y+}^{(0)}$, the zeroth order cross-ratio, is given in
\eq{ycross}. Here we have made the choice $r_3 =-1, r_2 =
\gamma_0, r_1 =\gamma_1, r_0=\gamma_\eta$. The cross-ratio at a
generic point near the $\Pi$-plane is given by \be{genpicrossp}
\eta_{\Pi+} = \eta_{\Pi+}^{(0)} - \frac{3}{16} \omega^2
\frac{2^{8/3} (s
\tilde\epsilon^3)^{2/3}}{(\gamma_0\gamma_1\gamma_\eta)^{1/6}
(\gamma_1 -\gamma_0)^{8/3} } \left( - \sqrt{ \gamma_0\gamma_1} ( 1
+\gamma_\eta) + \sqrt{\gamma_\eta} ( \gamma_0 + \gamma_1)
\right)^{4/3}, \ee where $ s\tilde \epsilon^3 = 1 + \gamma_1
-\gamma_\eta -\gamma_0$  and $\eta_{\Pi+}^{(0)}$ is the zeroth
order cross-ratio given in \eq{picross}. Here we have made the
choice
 $r_1 =\gamma_2, r_0=-\gamma_\eta, r_2 =-\gamma_1, r_3 =1$ to obtain the
 expansion about the $\Pi_2$ plane.
It is clear from the above expression that the perturbation
expansion of the Broom diagram about a generic point in the
$\Pi$-plane leads to a complex cross-ratio.

\section{The Square and the Whale diagrams}\label{squareapp}

\begin{figure}[htbp]
\begin{center}
\epsfig{file=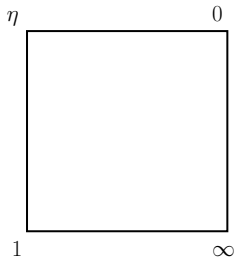, width=1.2in} \caption{The Square
diagram.}\label{Square}
\end{center}
\end{figure}
\begin{figure}[htbp]
\begin{center}
\vspace{.5in}
\epsfig{file=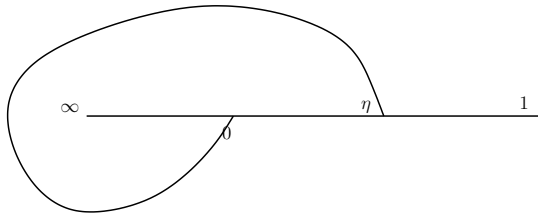, scale=0.5}
\vspace{.5in}
 \caption{The Whale diagram, which satisfies $\gamma_\eta+1=\gamma_0+\gamma_1$,
like the Square diagram.
 }\label{SquareSC}
\end{center}
\end{figure}

\noindent In this appendix we discuss two additional interesting
simple four-point function diagrams, the Square diagram (see
figure \ref{Square}) and the Whale diagram (see figure
\ref{SquareSC}). Both of them obey the constraint (defining as
before $\gamma_i=p_i/p_{\infty}$)
\be{SqCond}-\gamma_\eta+\gamma_0+\gamma_1=1.\ee
  The Strebel differential is given by \bea{StrebSq}
qdz^2=-p_\infty^2\frac{1}{4\pi^2}(d\ln F(z))^2,\qquad
F(z)=z^{\gamma_0}(z-\eta)^{-\gamma_\eta}(z-1)^{\gamma_1}. \eea The
poles of the differential are given by poles and zeros of $F$
(which are at $0,\eta,1$ and $\infty$). The zeros of the
differential are given by the zeros of $\d F$, which we will
denote by $c_\pm$. We have here one Strebel condition
\be{CondRe}{\rm Re}\left[\ln F(z)\right]|_{c_-}^{c_+}=0\qquad
\to\qquad FF^\dagger(c_+)=FF^\dagger(c_-).\ee Next, define \be{}
x=-\gamma_0\eta,\qquad y=\gamma_1(1-\eta),\ee so that the zeros
are given by: \bea{zerosSq}
2c_\pm=1-x-y\pm\sqrt{(1+x)^2+y^2-2y(1-x)}. \eea From here it's
easy to see that for $x$, $y$ real and $y_-\leq y\leq y_+$
($x_-\leq x\leq x_+$), where \be{} y_\pm=1-x\pm2\sqrt{-x}=(1\pm
\sqrt{-x})^2,\qquad (\,x_\pm=-1-y\pm2\sqrt{y}=-(1\mp
\sqrt{y})^2\,),\ee the reality condition \eqref{CondRe} is
satisfied as the two zeros are complex conjugates of each other.
Note that the zeros can be rewritten as: \bea{zerosSq2}
c_\pm=\left(\frac{\sqrt{y_+-y}\pm\sqrt{y_--y}}{2}\right)^2,\qquad
c_\pm-1=\left(\frac{\sqrt{x_+-x}\mp\sqrt{x_--x}}{2}\right)^2. \eea
Further, the dictionary between the two sets of variables is given
by \bea{dictSq} \eta=\frac{x(y-\bar y)}{y\bar x-x\bar y},\qquad
\gamma_0=\frac{x\bar y-\bar x y}{y-\bar y},\qquad
\gamma_1=\frac{x\bar y-\bar x y}{x-\bar x}. \eea Note that when
$\eta$ is real the transformation is singular, but as we saw above
in this case we can solve the Strebel problem.

\begin{figure}[htbp]
\begin{center}
$\begin{array}{c@{\hspace{1in}}c} \epsfig{file=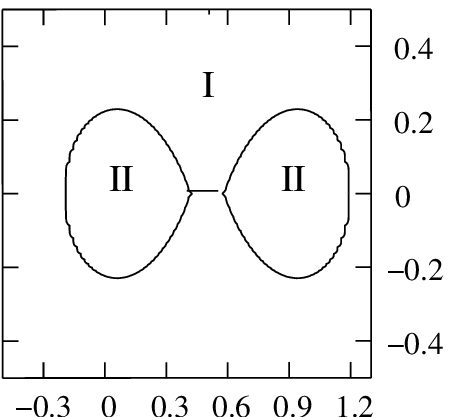, width=1.6in} &
    \epsfig{file=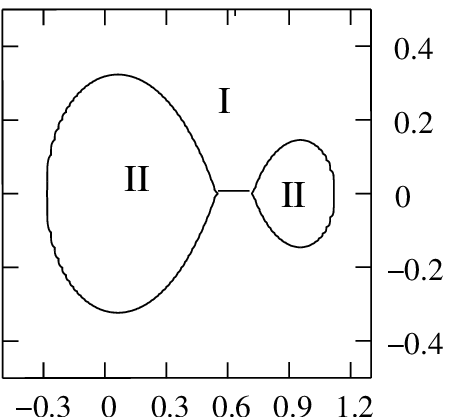, width=1.6in} \\ [0.4cm]
\end{array}$
\caption{ The curve on the complex $\eta$ plane for which the
Strebel condition is satisfied. On the left $\gamma_0=\gamma_1=70$
and on the right $\gamma_0=40$, $\gamma_1=70$. In general one can
see numerically that as $\gamma_0$ is taken to $1$ the curve
around $\eta=1$ shrinks to zero size, and as $\gamma_1$ is taken
to $1$ the curve around $\eta=0$ shrinks to zero size. This
implies that the diagrams corresponding to these curves cover the
full region corresponding to the OPE limits $\eta\to
0,1$.}\label{RegRe}
\end{center}
\end{figure}

From equation  (\ref{CondRe}) we see that there is no solution for
large $\eta$'s. Note that when we take $\eta$ to be large and at
least one of the circumferences does not scale to zero,
 one of the zeros $c_{\pm}$ becomes large
 and the other remains finite, and the condition \eq{CondRe} can not be
 satisfied.
Recall that $\g_0$ and $\g_1$ cannot both be small due to \eq{SqCond}. The fact that there are no solutions for
large $\eta$ means that the OPE of the vertices at $\eta$ and at $\infty$ is not covered by diagrams satisfying
(\ref{SqCond}). The solution of equation (\ref{CondRe}) can be obtained numerically, and has the general shape
depicted in figure \ref{RegRe}.
\noindent The nice feature of the Strebel differential
(\ref{StrebSq}) is that the horizontal leaves can be computed here
explicitly. The horizontal leaves satisfy \be
q\left(\frac{dz}{dt}\right)^2>0,\ee and thus \bea{leaves}
&&\frac{d}{dt}{\rm Re}[\ln F(z)]=0 \quad\to\quad
F(z(t))F^\dagger(\bar z(t))=C. \eea Here $C$ is a non-negative
constant parameterizing a certain leaf. The Strebel condition
(\ref{CondRe}) now has a nice interpretation: the $C$ parameter
for both zeros has to be equal, and this is required if we want a
leaf to go from one zero to another. Otherwise, either the leaves
emanating from a zero will not be compact (will not end) or they
will end on the same zero and the critical graph will be
disconnected. Both of these cases contradict the Strebel
conditions.

\begin{figure}[htbp]
\begin{center}
$\begin{array}{c@{\hspace{1in}}c} \epsfig{file=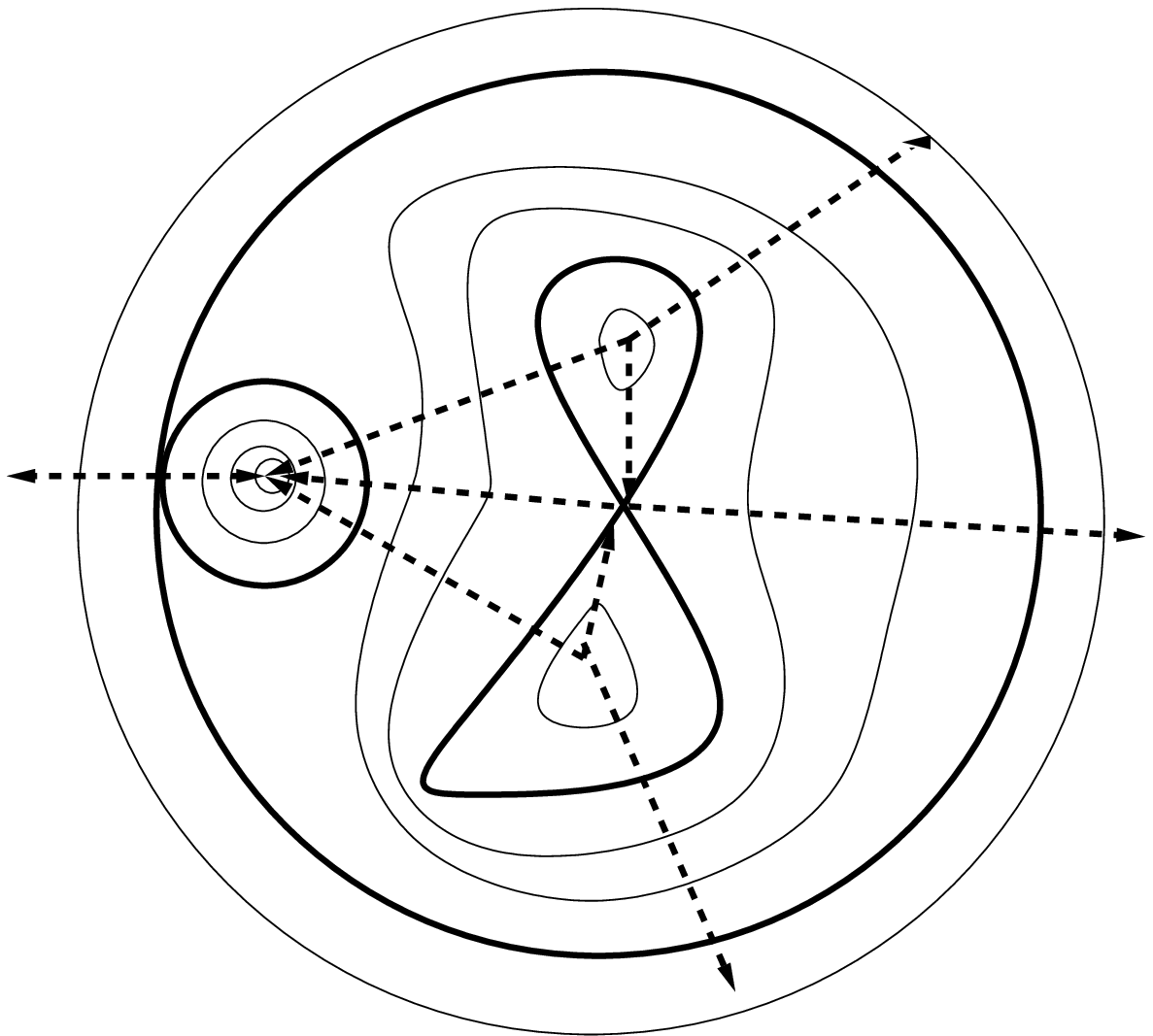, width=1.8in} &
    \epsfig{file=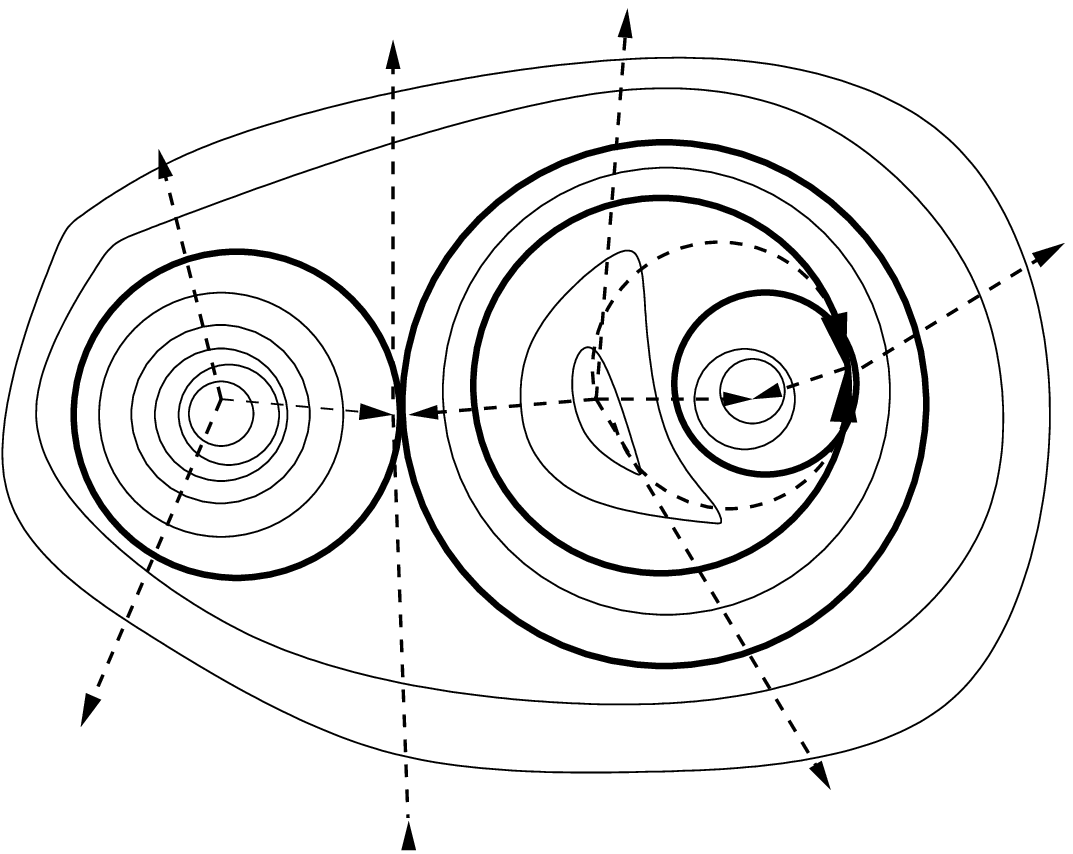, width=1.8in} \\ [0.4cm]
\end{array}$
\caption{The (disconnected) critical graphs and the leaves of the
two possible topologies. The arrows indicate directions of
increasing $C$. On the left we have shape I, and on the right
shape II.}\label{SqLeaves}
\end{center}
\end{figure}

We still have to identify which diagram the various solutions
depicted in figure \ref{RegRe} correspond to. The only two
diagrams which have the Strebel differential (\ref{StrebSq}) are
the Square diagram and the Whale diagram.
 We claim that the solution corresponding to the Square is the
real\footnote{ Note that these real solutions span only the
interval $\eta\in(0,1)$, and thus the cross-ratio of the square
can take values only in this interval.} $\eta$ solution on the
straight line in figure \ref{RegRe}, and the other solutions
correspond to the Whale. First, note that at a generic point of
the $\eta$ plane (where the Strebel condition is not satisfied)
the horizontal leaves begin and end on the same zero, and can have
only two topologically distinct shapes. This follows from the fact
that the nodes of the graph (the zeros of the Strebel
differential) correspond to saddle points of $FF^\dagger$ (as the
zeros are second order), and from the fact that the parameter $C$
goes to zero in the vicinity of $0$ and $1$ and diverges as one
approaches $\infty$ or $\eta$.  We can call the two shapes shape I
and shape II, see figure \ref{SqLeaves}. It is easy to show that
the graph in figure \ref{RegRe} is the boundary region between the
two different shapes. The real line is the boundary between two
different ways to get shape I, and the complex solution is the
boundary between shape I and II. When one goes from shape I to I
one gets a dual of a Square diagram, and when going from I to II
one obtains the dual of the Whale diagram. The topology in the
different regions is shown in figure \ref{RegRe}.

In the following figures we depict the critical graphs of the
Square, the Whale, and the additional diagrams that we get by
slightly deforming away from the solution to the Strebel condition
\eqref{CondRe}.
\begin{figure}[htbp]
\begin{center}
$\begin{array}{c@{\hspace{1in}}c} \epsfig{file=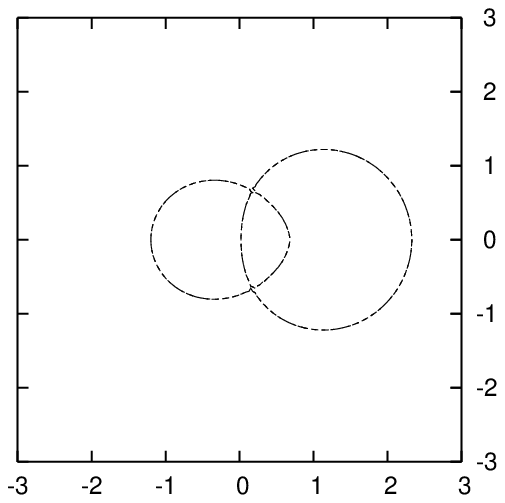, width=1.3in} &
    \epsfig{file=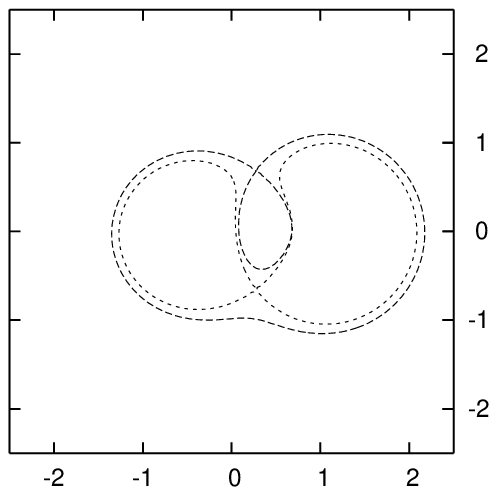, width=1.3in} \\ [0.4cm]
\end{array}$
\caption{On the left we draw the critical graph of the Square diagram for a typical real solution. On the right
we have the critical curves when slightly moving to the complex plane. We see how the graph smoothly becomes
disconnected, and has topology I.}\label{SqLeaves2}
\end{center}
\end{figure}

\begin{figure}[htbp]
\begin{center}
$\begin{array}{c@{\hspace{.2in}}c@{\hspace{.2in}}c@{\hspace{.2in}}c}
\epsfig{file=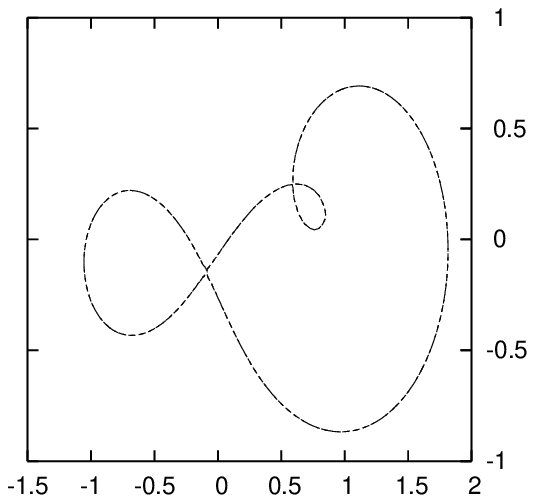, width=1.3in} &
    \epsfig{file=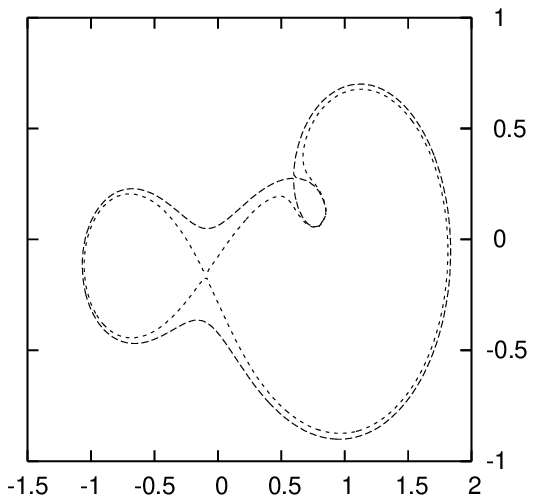, width=1.3in} &
\epsfig{file=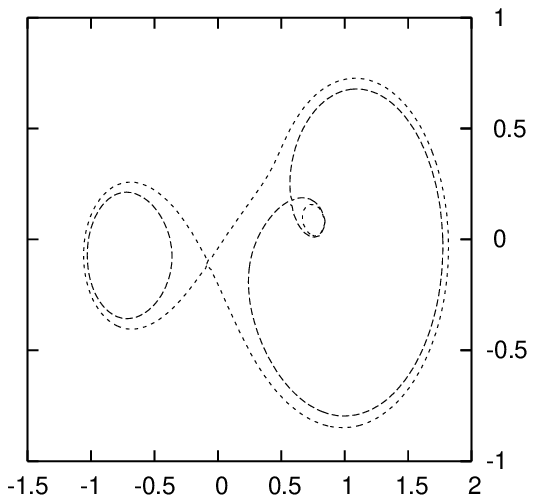, width=1.3in} &
\epsfig{file=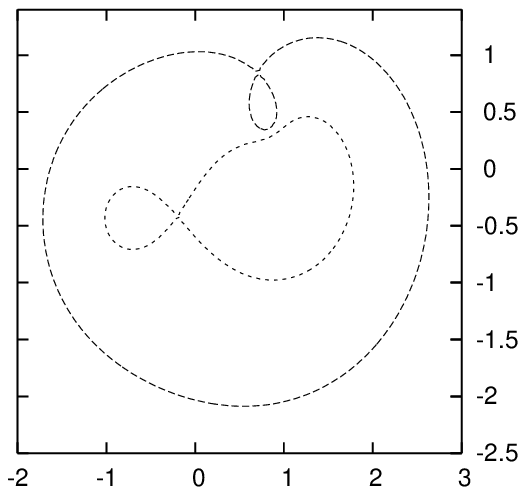, width=1.3in}
\\ [0.4cm]
\end{array}$
\caption{On the left, we draw the critical graph of a
complex solution for $\eta$. It is easy to see that this critical
graph is dual to the Whale diagram. In the next two graphs we
again slightly move away from the point where the Strebel
conditions are satisfied. We keep the same real part of $\eta$ and
first move up and then down along the imaginary axis. The critical
graph has topology II when lowering ${\rm Im}(\eta)$, and topology
I when raising ${\rm Im}(\eta)$. The last graph is a generic
complex $\eta$ not satisfying the Strebel conditions. The critical
graph is disconnected and we do not have a cell decomposition.
}\label{NotSqLeaves}
\end{center}
\end{figure}


\newpage
\begin{thebibliography}{10}

\bibitem{Gopakumars}
R.~Gopakumar, ``From free fields to AdS,'' Phys.\ Rev.\ D {\bf 70}, 025009 (2004) [arXiv:hep-th/0308184];
R.~Gopakumar, ``From free fields to AdS. II,'' Phys.\ Rev.\ D {\bf 70}, 025010 (2004) [arXiv:hep-th/0402063];
R.~Gopakumar, ``Free field theory as a string theory?,'' Comptes Rendus Physique {\bf 5}, 1111 (2004)
[arXiv:hep-th/0409233]; R.~Gopakumar, ``From free fields to AdS. III,'' Phys.\ Rev.\ D {\bf 72}, 066008 (2005)
[arXiv:hep-th/0504229].

\bibitem{'tHooft:1973jz}
G.~'t Hooft, ``A planar diagram theory for strong interactions,''
Nucl.\ Phys.\ B {\bf 72}, 461 (1974).

\bibitem{GopakumarVafa}
  R.~Gopakumar and C.~Vafa,
  ``On the gauge theory/geometry correspondence,''
  Adv.\ Theor.\ Math.\ Phys.\  {\bf 3}, 1415 (1999)
  [arXiv:hep-th/9811131];
  H.~Ooguri and C.~Vafa,
  ``Worldsheet derivation of a large N duality,''
  Nucl.\ Phys.\ B {\bf 641}, 3 (2002)
  [arXiv:hep-th/0205297].

\bibitem{Konts}
E.~Witten,
  ``Two-dimensional gravity and intersection theory on moduli space,''
  Surveys Diff.\ Geom.\  {\bf 1}, 243 (1991);
  M.~Kontsevich,
  ``Intersection theory on the moduli space of curves and the matrix Airy
  function,''
  Commun.\ Math.\ Phys.\  {\bf 147}, 1 (1992);
   D.~Gaiotto and L.~Rastelli,
  ``A paradigm of open/closed duality: Liouville D-branes and the  Kontsevich
  model,''
  JHEP {\bf 0507}, 053 (2005)
  [arXiv:hep-th/0312196].

\bibitem{Maldacena:1997re}
  J.~M.~Maldacena,
  ``The large N limit of superconformal field theories and supergravity,''
  Adv.\ Theor.\ Math.\ Phys.\  {\bf 2}, 231 (1998)
  [Int.\ J.\ Theor.\ Phys.\  {\bf 38}, 1113 (1999)]
  [arXiv:hep-th/9711200];
  S.~S.~Gubser, I.~R.~Klebanov and A.~M.~Polyakov,
  ``Gauge theory correlators from non-critical string theory,''
  Phys.\ Lett.\ B {\bf 428} (1998) 105
  [arXiv:hep-th/9802109];
  E.~Witten,
  ``Anti-de Sitter space and holography,''
  Adv.\ Theor.\ Math.\ Phys.\  {\bf 2}, 253 (1998)
  [arXiv:hep-th/9802150].


\bibitem{FreeFields}
K.~Bardakci and C.~B.~Thorn,
``A worldsheet description of large $N_c$ quantum field theory,''
Nucl.\ Phys.\ B {\bf 626}, 287 (2002)
[arXiv:hep-th/0110301]; C.~B.~Thorn,
``A worldsheet description of planar Yang-Mills theory,''
Nucl.\ Phys.\ B {\bf 637}, 272 (2002)
[Erratum-ibid.\ B {\bf 648}, 457 (2003)]
[arXiv:hep-th/0203167]; K.~Bardakci and C.~B.~Thorn,
``A mean field approximation to the world sheet model of planar $\phi^3$ field
theory,''
Nucl.\ Phys.\ B {\bf 652}, 196 (2003)
[arXiv:hep-th/0206205]; S.~Gudmundsson, C.~B.~Thorn and T.~A.~Tran,
``BT worldsheet for supersymmetric gauge theories,''
Nucl.\ Phys.\ B {\bf 649}, 3 (2003)
[arXiv:hep-th/0209102]; K.~Bardakci and C.~B.~Thorn,
``An improved mean field approximation on the worldsheet for planar $\phi^3$
theory,''
Nucl.\ Phys.\ B {\bf 661}, 235 (2003)
[arXiv:hep-th/0212254]; C.~B.~Thorn and T.~A.~Tran,
``The fishnet as anti-ferromagnetic phase of worldsheet Ising spins,''
Nucl.\ Phys.\ B {\bf 677}, 289 (2004)
[arXiv:hep-th/0307203];
 K.~Bardakci,
  ``Further results about field theory on the world sheet and string
  formation,''
  Nucl.\ Phys.\ B {\bf 715} (2005) 141
  [arXiv:hep-th/0501107];
  M.~Kruczenski,
  ``Planar diagrams in light-cone gauge,''
  JHEP {\bf 0610}, 085 (2006)
  [arXiv:hep-th/0603202];
K.~Bardakci,
  ``Field theory on the world sheet: Mean field expansion and cutoff
  dependence,''
  [arXiv:hep-th/0701098].

\bibitem{StringBits}
P.~Haggi-Mani and B.~Sundborg,
  ``Free large N supersymmetric Yang-Mills theory as a string theory,''
  JHEP {\bf 0004}, 031 (2000)
  [arXiv:hep-th/0002189];
  H.~L.~Verlinde,
  ``Bits, matrices and $1/N$,''
  JHEP {\bf 0312} (2003) 052
  [arXiv:hep-th/0206059];
  J.~G.~Zhou,
  ``pp-wave string interactions from string bit model,''
  Phys.\ Rev.\ D {\bf 67} (2003) 026010
  [arXiv:hep-th/0208232];
  D.~Vaman and H.~L.~Verlinde,
  ``Bit strings from N = 4 gauge theory,''
  JHEP {\bf 0311} (2003) 041
  [arXiv:hep-th/0209215];
  A.~Dhar, G.~Mandal and S.~R.~Wadia,
  ``String bits in small radius AdS and weakly coupled N = 4 super  Yang-Mills
  theory. I,''
  arXiv:hep-th/0304062;
  K.~Okuyama and L.~S.~Tseng,
  ``Three-point functions in N = 4 SYM theory at one-loop,''
  JHEP {\bf 0408} (2004) 055 [arXiv:hep-th/0404190];
  L.~F.~Alday, J.~R.~David, E.~Gava and K.~S.~Narain,
  ``Structure constants of planar N = 4 Yang Mills at one loop,''
  JHEP {\bf 0509} (2005) 070
  [arXiv:hep-th/0502186];
  J.~Engquist and P.~Sundell,
  ``Brane partons and singleton strings,''
  arXiv:hep-th/0508124;
  L.~F.~Alday, J.~R.~David, E.~Gava and K.~S.~Narain,
  ``Towards a string bit formulation of N = 4 super Yang-Mills,''
  arXiv:hep-th/0510264.

\bibitem{Joe}
J.~Polchinski, unpublished.

\bibitem{Karch}
A.~Karch,
``Lightcone quantization of string theory duals of free field
theories,'' arXiv:hep-th/0212041;
A.~Clark, A.~Karch, P.~Kovtun and D.~Yamada,
``Construction of bosonic string theory on infinitely curved
anti-de  Sitter space,'' Phys.\ Rev.\ D {\bf 68}, 066011 (2003)
[arXiv:hep-th/0304107].

\bibitem{Bonelli:2004ve}
G.~Bonelli, ``On the boundary gauge dual of closed tensionless
free strings in AdS,'' JHEP {\bf 0411} (2004) 059
[arXiv:hep-th/0407144].

\bibitem{Itzhaki:2004te}
  N.~Itzhaki and J.~McGreevy,
  ``The large N harmonic oscillator as a string theory,''
  Phys.\ Rev.\ D {\bf 71} (2005) 025003
  [arXiv:hep-th/0408180].

\bibitem{Bianchi}
  M.~Bianchi, J.~F.~Morales and H.~Samtleben,
  ``On stringy AdS(5) x S**5 and higher spin holography,''
  JHEP {\bf 0307} (2003) 062
  [arXiv:hep-th/0305052];
  N.~Beisert, M.~Bianchi, J.~F.~Morales and H.~Samtleben,
  ``On the spectrum of AdS/CFT beyond supergravity,''
  JHEP {\bf 0402} (2004) 001
  [arXiv:hep-th/0310292];
  D.~E.~Diaz and H.~Dorn,
  ``On the AdS higher spin / O(N) vector model correspondence: Degeneracy of
  the holographic image,''
  JHEP {\bf 0607}, 022 (2006)
  [arXiv:hep-th/0603084].


\bibitem{Akhmedov}
  E.~T.~Akhmedov,
  ``Expansion in Feynman graphs as simplicial string theory,''
  JETP Lett.\  {\bf 80}, 218 (2004)
  [Pisma Zh.\ Eksp.\ Teor.\ Fiz.\  {\bf 80}, 247 (2004)]
  [arXiv:hep-th/0407018].


\bibitem{Carfora:2006nj}
  M.~Carfora, C.~Dappiaggi and V.~L.~Gili,
  ``Triangulated surfaces in twistor space: A kinematical set up for open /
  closed string duality,''
  JHEP {\bf 0612}, 017 (2006)
  [arXiv:hep-th/0607146];
  M.~Carfora, C.~Dappiaggi and V.~L.~Gili,
  ``From random Regge triangulations to open strings,''
  arXiv:hep-th/0702114.


\bibitem{Furuuchi:2005qm}
  K.~Furuuchi,
  ``From free fields to AdS: Thermal case,''
  Phys.\ Rev.\ D {\bf 72} (2005) 066009
  [arXiv:hep-th/0505148].

\bibitem{Aharony:2006th}
O.~Aharony, Z.~Komargodski and S.~S.~Razamat, ``On the worldsheet
theories of strings dual to free large N gauge theories,'' JHEP
{\bf 0605}, 016 (2006) [arXiv:hep-th/0602226].

\bibitem{David:2006qc}
  J.~R.~David and R.~Gopakumar,
  ``From spacetime to worldsheet: Four point correlators,''
  JHEP {\bf 0701}, 063 (2007)
  [arXiv:hep-th/0606078].


\bibitem{Yaakov:2006ce}
  I.~Yaakov,
  ``Open and closed string worldsheets from free large N gauge theories with
  adjoint and fundamental matter,''
  JHEP {\bf 0611}, 065 (2006)
  [arXiv:hep-th/0607244].

\bibitem{K.Strebel:1984}
K.~Strebel, ``Quadratic differentials,'' Springer-Verlag, 1984.

\bibitem{Mulase:98}
M.~Mulase, M.~Penkava, ``Ribbon graphs, quadratic differentials on
Riemann surfaces, and algebraic curves defined over $\bar Q$ ,''
[math-ph/9811024].

\bibitem{Zvonkine:2002}
D.~Zvonkine, ``Strebel differentials on stable curves and
Kontsevich's proof of Witten's conjecture,'' [math.AG/0209071].



\bibitem{Zwiebach}
B.~Zwiebach,
  ``Closed string field theory: Quantum action and the B-V master equation,''
  Nucl.\ Phys.\ B {\bf 390}, 33 (1993)
  [arXiv:hep-th/9206084];
A.~Belopolsky and B.~Zwiebach,
  ``Off-shell closed string amplitudes: Towards a computation of the tachyon
  potential,''
  Nucl.\ Phys.\ B {\bf 442}, 494 (1995)
  [arXiv:hep-th/9409015].

\bibitem{Moeller:2004yy}
N.~Moeller, ``Closed bosonic string field theory at quartic
order,'' JHEP {\bf 0411} (2004) 018 [arXiv:hep-th/0408067];
  N.~Moeller,
  ``Closed bosonic string field theory at quintic order: Five-tachyon contact
  term and dilaton theorem,''
  arXiv:hep-th/0609209.

\bibitem{Ashok:2006du}
  S.~K.~Ashok, F.~Cachazo and E.~Dell'Aquila,
  ``Strebel differentials with integral lengths and Argyres-Douglas
  singularities,''
  arXiv:hep-th/0610080.


\bibitem{Gross:1987ar}
D.~J.~Gross and P.~F.~Mende,
  ``The High-Energy Behavior Of String Scattering Amplitudes,''
  Phys.\ Lett.\  B {\bf 197}, 129 (1987);
  D.~J.~Gross and P.~F.~Mende,
  ``String Theory Beyond The Planck Scale,''
  Nucl.\ Phys.\ B {\bf 303}, 407 (1988).

\bibitem{Brower:2006ea}
  R.~C.~Brower, J.~Polchinski, M.~J.~Strassler and C.~I.~Tan,
  ``The pomeron and gauge / string duality,''
  arXiv:hep-th/0603115.

\bibitem{ads3cft2}
  J.~M.~Maldacena and H.~Ooguri,
  ``Strings in AdS(3) and the SL(2,R) WZW model. III: Correlation  functions,''
  Phys.\ Rev.\  D {\bf 65}, 106006 (2002)
  [arXiv:hep-th/0111180];
  J.~R.~David, G.~Mandal and S.~R.~Wadia,
  ``Microscopic formulation of black holes in string theory,''
  Phys.\ Rept.\  {\bf 369}, 549 (2002)
  [arXiv:hep-th/0203048];
  M.~R.~Gaberdiel and I.~Kirsch,
  ``Worldsheet correlators in AdS(3)/CFT(2),''
  arXiv:hep-th/0703001;
    A.~Dabholkar and A.~Pakman,
  ``Exact chiral ring of AdS(3)/CFT(2),''
  arXiv:hep-th/0703022.

\end{thebibliography}
\end{document}